\documentclass[useAMS,usenatbib]{mn2e}
\usepackage{graphicx,color,epsfig}

\def\ltsima{$\; \buildrel < \over \sim \;$}
\def\simlt{\lower.5ex\hbox{\ltsima}}
\def\gtsima{$\; \buildrel > \over \sim \;$}
\def\simgt{\lower.5ex\hbox{\gtsima}}

\newcommand\mearth{{\,{\rm M}_{\oplus}}}
\newcommand{\apj}{ApJ}

\newcommand{\mnras}{MNRAS}
\newcommand{\aap}{A\&A}
\newcommand{\araa}{ARA\&A}
\newcommand{\apjl}{ApJL}
\newcommand{\aj}{AJ}
\newcommand{\nat}{Nature}

\newcommand{\icarus}{ICARUS}

\title[Origin of planetesimals]{An alternative origin for debris rings of
  planetesimals.}

\author[S. Nayakshin \& S.-H. Cha] {Sergei Nayakshin and Seung-Hoon Cha
\\ Department of Physics \&
  Astronomy, University of Leicester, Leicester, LE1 7RH, UK}

\begin{document}

\date{Received}

\pagerange{\pageref{firstpage}--\pageref{lastpage}} \pubyear{2011}

\maketitle

\label{firstpage}

\begin{abstract}
Core Accretion (CA), the most widely accepted scenario for planet formation,
postulates existence of $\sim$ km-sized solid bodies, called planetesimals,
arranged in a razor-thin disc in the earliest phases of planet
formation. These objects coagulate by collisions, eventually building
planetary cores. In the Tidal Downsizing (TD) hypothesis, an alternative
scenario for formation of planets, grain growth, sedimentation and formation
of planetary cores occur inside dense and massive gas clumps formed in the
outer cold disc by gravitational instability. As a clump migrates inward,
tidal forces of the star remove all or most of the gas from the clump,
downsizing it to a planetary mass body. Here we consider a rotating and/or
strongly convective gas clump. We argue that such a clump may form not
only the planetary core but also numerous smaller bodies. As an example, we
consider the simplest case of bodies on circular orbits around the
planetary core in the centre of the gas clump. We find that bodies smaller
than $\sim 1$ km suffer a strong enough aerodynamic drag and thus spiral in
and accrete onto the solid core rapidly. On the contrary, bodies in the
planetesimal and larger size range lose their centrifugal support very
slowly. We consider analytically and numerically the fate of these bodies
after the host gas clump is disrupted. Planetesimals orbiting the
protoplanetary core closely remain gravitationally bound to it; these may be
relevant to formation of satellites of giant planets. Planetesimals on more
distant orbits within the host clump are unbound from the protoplanet and are
set on mildly eccentric heliocentric orbits, generically forming wide
rings. These may correspond to debris discs around main sequence stars and the
Kuiper belt in the Solar System. For the latter in particular, TD hypothesis
naturally explains the observed sharp outer edge and the ``mass deficit'' of
the Kuiper belt.
\end{abstract}


\section{Introduction}

\subsection{Planet formation: bottom-up or top-down?}

Core Accretion (CA) theory \citep{PollackEtal96,AlibertEtal05} posits that
planet formation starts when some of the dust grains in the protoplanetary
disc grow and sediment to the disc midplane.  This step is then followed by a
less well understood one in which much larger solid objects of a few km or
more are made \citep[cf.][]{Youdin02,JohansenEtal07}. These rocky objects,
called ``planetesimals'', then co-coagulate into terrestrial-like planetary
cores \citep{Safronov72}. These cores continue to gain mass by mergers and
accretion of planetesimals.  If the cores are still embedded in the parent
gaseous disc, then a gas atmosphere builds up around the cores. When the solid
core mass exceeds a few to a few tens Earth masses, the atmosphere becomes
self-gravitating and collapses hydrodynamically to much higher densities,
forming a proto-giant gas planet \citep{Mizuno80,Stevenson82,Rafikov06}. This
model may be termed a bottom-up scenario for planet formation.

An alternative top-down scenario for the origin of planets is also physically
plausible, although it is currently much less developed
\citep{BoleyEtal10,Nayakshin10b}. The very first step in this ``Tidal
Downsizing'' (TD) scheme is reminiscent of formation of stars, that is the
Jeans self-gravitation instability of a gas cloud/clump \citep[for example,
  see][]{Larson69}.  The cloud is however much less massive, e.g., its mass is
around the opacity fragmentation limit of $\sim 10 M_J$
\citep{Rees76,Nayakshin10a,ForganRice11}, and is born inside a gas disc
orbiting the parent star \citep[e.g.,][]{Boss97}. The second step in TD
scenario is similar to the first one in CA model, e.g., grains grow and
sediment. However, this process occurs inside the gas clumps rather than the
whole of the disc, where the densities are much lower \citep[cf. simulations
  of][]{ChaNayakshin11a}. A massive solid core therefore forms inside the gas
clump \citep{Boss98,HS08,Nayakshin10b}. Rapid inward radial migration of the
gas clump \citep{VB05,VB06,MachidaEtal11,MichaelEtal11,BaruteauEtal11} then
expose the clump to the ever increasing tidal force of the parent star.
Removal of all or a part of the original gas cloud by tidal forces of the star
may hypothetically leave behind both terrestrial-like and gas giant planets
\citep{BoleyEtal10,Nayakshin10c}. Note that while the combination of these
physical steps is a very recent development, the TD hypothesis may be viewed
as a physics-upgraded version of the gravitational disc instability model
\citep[GI; e.g.,][]{Boss97,Rice05,Rafikov05}. The complicated fate of gas
clumps in a self-gravitating disc was also discussed by \cite{MayerEtal04},
who also noted that these clumps may be tidally disrupted if they migrate
inward. Furthermore, except for the crucial step of the radial
migration\footnote{\cite{DW75} showed that the gas clump with properties
  envisaged by \cite{McCreaWilliams65}, sitting at the present location of the
  Earth, would be tidally disrupted well before grains could have grown and
  sedimented to its centre. Making the clump at $\sim 100$ AU, where the clump
  may be much cooler, and having grains sediment into a massive solid core
  before bringing the clump into the inner Solar System by migration is the
  only physically plausible way for the model to work.} of the gas clump, the
TD scheme is similar to ideas of \cite{McCreaWilliams65}.

The current status of TD planet formation hypothesis is best described as
``work in progress''. Progress is being made in terms of building more
self-consistent time-dependent models of the protoplanetary discs with
embedded massive gas clumps \citep[e.g.,][]{BoleyEtal11a,NayakshinLodato11};
addressing the surprising high temperature content of Solar System comets
\citep{Vorobyov11a,NayakshinEtal11a}; chemistry of self-gravitating discs
\citep{IleeEtal11}; the ``hot'' super-Earth planets \citep{Nayakshin11b}; and
rotation of the Solar System terrestrial planets \citep{Nayakshin11a}.
Despite this, no detailed statistical predictions in the spirit of population
synthesis models of \cite{IdaLin08,IdaLin10} have yet been made. The TD
hypothesis thus cannot yet be compared with exoplanetary data in detail.

One promising way to constrain the TD hypothesis observationally is via
observations of the early ``embedded'' phase of star (and potentially planet)
formation \citep{DunhamVorobyov12}, especially with the advent of the {\em
  ALMA} telescope when resolving individual massive gas clumps in large $R
\simgt 50$ AU discs around young protostars may become possible.

\subsection{Can solid debris constrain planet formation theories?}\label{sec:debris_intro}

Planets are not the only bodies orbiting their host stars: solid bodies from
microscopic dust to sub-terrestrial planet size objects such as Pluto also
form during planet assembly both in the Solar System and around other nearby
stars. In particular, Solar System contains several repositories of $\simgt 1$
km sized solid bodies -- the asteroid belt, the Kuiper belt, the scattered
disc and the Oort cloud of long-period comets.  Starting from the discovery of
the infrared excess beyond 12$\mu$m around Vega by IRAS \citep{Aumann84},
circum-stellar dust disc are commonly found around pre- and main sequence stars
\citep{Zuckerman01,Wyatt08}. Since microscopic dust in the discs around main
sequence dust should be blown away rapidly due to radiation pressure from the
parent stars, a continuous replenishment source for the dust is required
\citep{WyattEtal07}. The universally accepted picture is that the
circum-stellar dust results from a fragmentation "cascade"
\citep[e.g.,][]{Hellyer70,WyattEtal07b,Wyatt08} of larger solid bodies such as
comets and asteroids. This view is reinforced by observations of dust
resulting from such fragmentation cascades in our own Solar System
\citep{NesvornyEtal03,NesvornyEtal10}, and also by the signatures of
collisional sculpting of the asteroid belt \citep[e.g.,][]{BottkeEtal05}.

It is obvious that the solid {\em debris} around stars has a rather natural
explanation in the context of the CA theory: these bodies are the
planetesimals and ``half-grown'' planetary embryos that did not get consumed
or ejected completely from the system by the growing planets. The question we
pose in this paper is this: Can TD hypothesis produce solid debris too, and if
so, how do the properties of that debris differ from the CA model?

We shall argue below that it is economical and logical to have all smaller
solid objects to be born inside the same ``parent'' gas clumps where the
terrestrial-like solid cores grow.  We arrive at these conclusions based on
the following ideas. In the 1D spherically symmetric models of Helled and
co-authors, and in \cite{Nayakshin10b}, only one central core is
formed. However, \citep{Nayakshin11b} has shown that specific angular momentum
of the grains may be too large to allow gravitational collapse into just one
massive body. Gravitational collapse of a rapidly rotating cloud may result in
formation of not only the central body but also a number of smaller objects
possibly orbiting the larger one in a centrifugally supporting disc. The
objects may fragment or grow further due to subsequent collisions. In
addition, the formation of the solid core releases a significant amount of
energy into the surrounding gas \citep{Nayakshin10b}, which stirs up strong
convective motions. This second effect may also promote formation of
additional solid bodies by gravitational collapse of smaller grain-dominated
regions \citep[cf. \S 3.6.2 and 3.6.3 of][]{Nayakshin10a} on chaotically
oriented orbits.

The next logical step in our picture is the disruption of the parent gas
clump, as required by the TD hypothesis. We show below that the solid debris
population within the clump can be divided into two groups: one bound to the
planetary core (be it terrestrial-like or more massive by that point), and the
other to the host gas clump. Upon disruption of the clump the first group
``survives'' and becomes planet's satellites; the other group gets disrupted
and dispersed with the gas. Generically, the second population of debris forms
a ``disruption ring'' centred on the location of the host gas clump
disruption.

Below we study analytically and numerically (for three representative cases)
the properties of the post-disruption orbits of the planetesimal debris. We
start in \S 2 by discussing the likely structure of the gas clump before the
disruption. We show that there are two possible ways of the host clump
disruption -- either tidally or due to an internal energy release by the
growing protoplanetary core itself. We estimate the minimum mass of the core
being able to disrupt the host clump to be $\sim 10$ Earth masses. 

In \S \ref{sec:drag} we consider the aerodynamic drag acting on the solid
bodies within the host clump. We find that bodies smaller than $\sim$ 1~cm and
larger than $\sim 1$~km stand a good chance of remaining ``independent'' after
the clump disruption compared with the bodies between these size limits: these
suffer strong drag from the gas and must end up joining the protoplanetary
core in the centre of the gas clump. In \S \ref{sec:theory} we consider
analytically the kind of orbits that the large solid bodies obtain after the
gas clump disruption. \S \ref{sec:numerics} describes the set up of our
numerical experiments. Several following sections present the numerical
results, with \S \ref{sec:gas} showing the gas flow, \S\S \ref{sec:unbound}
and \ref{sec:bound_orbits} focusing on the population of unbound and bound
solid bodies, respectively.  \S \ref{sec:discussion} contains a discussion of
the main results of our paper.

\section{Host clump structure and disruption}\label{sec:host}

\subsection{Setup and terminology}\label{sec:defs}

For definitiveness, we consider a host gas clump of $M_{\rm hc} = 5 M_J$ mass
in the numerical part of the paper. We assume that the initial density and
temperature profiles are that of a polytropic sphere with the polytropic index
$n=5/2$ as appropriate for a molecular hydrogen-dominated gas in the
temperature range from a few hundred K to about 1500 K
\citep{BoleyEtal07}. The clump is initially located at $R=40$ AU on a circular
orbit around the star of mass $M_* = 1 M_\odot$.

We shall refer to the host gas clump as such to distinguish it from the solid
planetary core, which we frequently call ``planet''. The planet (of mass
$M_{\rm p} = 10 \mearth$) is treated as a point mass in this paper. We assume
that the density of the planet is much higher than that of the host
clump. Therefore, the internal structure of the planet is not affected by the
host gas clump disruption. We note that this setting does not require the
planet to consist of high-Z material only; volatiles could be present also as
long as their densities are much higher than the tidal density, $M_*/2\pi
R^3$.

\subsection{Clump disruption by tidal forces}\label{sec:tidal}

We study two limiting cases of clump disruption. In the case of disruption due
to tidal forces from the star, the host clump fills its Roche lobe at the
start of the simulation, time $t=0$. The size of the Roche lobe (Hill's
radius) at this location is
\begin{equation}
\label{eq:hillsr}
r_h = R\left(\frac{M_{\rm hc}}{3M_*}\right)^{1/3}\approx 4.7\;\mbox{AU}\;.
\end{equation}
A polytropic cloud with the clump mass $M_{\rm hc} = 5 M_J$ and the clump
radius, $r_{\rm hc}$, satisfying $r_{\rm hc} = r_h$ has central temperature $T
= 195$ K, for reference.

The mass-radius relation of a polytropic cloud is given by
\begin{equation}
r_{\rm hc} \propto M_{\rm hc}^\frac{1-n}{3-n} \propto M_{\rm hc}^{-3} \mbox{
  for } n = 5/2\;.
\label{mass_radius}
\end{equation}
This implies that the host clump expands rapidly as mass is lost. We therefore
expect a prompt tidal disruption of such a host clump by Roche lobe overflow:
$r_{\rm hc}$ increases while $r_h$ decreases as $M_{\rm hc}$ drops.

We shall add to this that the stabilising effect of the host clump outward
migration due to the mass exchange between the gas clump and the star,
discovered by \cite{NayakshinLodato11}, does not occur in our simulations
because the clump is destroyed rather rapidly in all of the simulations below.
In addition, we find that disruption proceeds via both L1 and L2 points, in
contrast to what is found for much more gentle ``hot disruptions'' at $R \sim
0.1$ AU separations by \cite{NayakshinLodato11}. For both of these reasons
there is no outward torque on the host clump; the tidal disruption has a
runaway character.

\subsection{Disruption by an internal energy release}\label{sec:explosion}

In the second class of models we consider here, the clump's initial radius is
smaller than the Hill's radius $r_h$. In particular, for the two simulations
presented below, we set the initial central temperature to $T=500$ K, which
corresponds to the clump's initial radius of $r_{\rm hc}=1.84$ AU. We then
assume that a sudden burst of energy release occurs in the centre of the
clump. Physically, we relate the burst to the assembly of the planetary core,
as is found in 1D simulations of \cite{Nayakshin10b}.  In detail, the energy
released by the solid core is passed to the surrounding gas by radiative
diffusion. If the radiative diffusion time scale is longer than the planetary
core's assembly time and the injection energy is large enough then even an
isolated gas clump may be disrupted. An example of this is simulation
M0$\alpha$3, Figures 5 to 7 in \cite{Nayakshin10b}.

We estimate the energy
released by the planetary core to be about its binding energy,
\begin{equation}
E_{\rm bind, c} \approx\frac{G M_p^2}{2 r_p} = 5 \times 10^{40} \;
\hbox{erg}\; \left(\frac{M_p}{10 M_\oplus}\right)^{5/3} \rho_p^{1/3} \;,
\label{ebind_p}
\end{equation}
where $G$ is the gravitational constant, and $\rho_p$ is the solid core
density in g cm$^{-3}$.

Let us now compare the core's binding energy with the minimum amount of energy
needed to disrupt the host clump.  The total energy of a polytropic gas clump,
$E_{\rm hc}$, is given by
\begin{equation}
\label{eq:tot_poly}
E_{\rm hc} = -\frac{3-n}{5-n}\frac{GM_{\rm hc}^2}{r_{\rm hc}}\;.
\end{equation}
For simplicity we assume that the clump remains polytropic after the energy
injection (which is not obvious at all; we will come back to this later) with
same $n$ but a different adiabatic constant. The energy input sufficient to
disrupt the clump, $\Delta U_0$ is the energy needed to inflate the clump to
the point of its tidal disruption, i.e., when $r_{\rm hc}$ increases and
becomes equal to the Hill's radius, $r_h$. Thus,
\begin{equation}
\label{eq:ene_dis}
\Delta U_0 = \frac{3-n}{5-n} \left(1-\frac{r_{\rm hc}}{r_h}\right)\;
\frac{GM_{\rm hc}^2}{r_{\rm hc}}\;.
\end{equation}
For reference, $GM_{\rm hc}^2/r_{\rm hc}\approx 2.4\times 10^{41} $erg for
clump mass $M_{\rm hc} = 5 M_J$ and clump radius $r_{\rm hc} = 1.84$ AU.

As energy released by the core heats the surrounding gas, it is instructive to
compare the required energy injection $\Delta U_0$ with the initial total
internal energy of the host clump, $U_0$:
\begin{equation}
\label{deltau_disr}
\frac{\Delta U_0}{U_0} =
-\left(1-\frac{3}{n}\right)\left(1-\frac{r_{\rm hc}}{r_h}\right)\;.
\end{equation}
In the limit of the initially small host clump, $r_{\rm hc} \ll r_h$, the
disruption energy is $\Delta U_0 = 0.2 U_0$ for $n=5/2$. In the case under
consideration, $r_{\rm hc} = 1.84$ AU, and $r_h=4.7$ AU, and thus $\Delta U_0
= 0.123 U_0$.

From the following we conclude that the energy required to unbind the host gas
clump is of the order of $3\times 10^{40}$ erg. This is of the same order as
the binding energy of the planetary core of mass $M_p=10 M_\oplus$. Therefore,
we see that, to disrupt the host clump with the energy released by the
planetary core, the mass of the core must be at least $M_p \sim 10
M_\oplus$. We also note in passing that the similarity of this mass to the
critical solid core mass in the CA theory
\citep[e.g.,][]{Mizuno80,Stevenson82} seems to be a pure coincidence. While
the minimum mass capable of disrupting the gas host clump is the function of
the conditions in that clump (mass, age, opacity, etc.), the CA critical mass
is a function of the core's surroundings -- the location from the star,
opacity and the planetesimal accretion rate from the surrounding
protoplanetary disc \citep[e.g.,][]{Rafikov11}.

\section{Solids within the clump}\label{sec:solids_within}

\cite{Nayakshin10a,Nayakshin10b} finds that when grains contained in the host
clump grow by the hit-and-stick mechanism to sizes of the order of $s\sim 10$
cm, their sedimentation within $\sim 1000$ years becomes possible \citep[see
  also][]{McCreaWilliams65,Boss98}. Accumulation of these to the centre of the
clump creates a region dominated by grains rather than by gas. This region,
called ``grain cluster'' by \cite{Nayakshin10a} then becomes gravitationally
unstable and collapses to form a massive core composed of high-Z materials.

As argued in \S \ref{sec:debris_intro}, due to rotation and chaotic convective
gas motions in the centre of the host clump, one may expect numerous smaller
bodies to form in the centre of the clump as well. It is the fate of these
bodies that interests us here.

\subsection{Aerodynamic drag in the host clump}\label{sec:drag}

We shall now consider what happens to solids of different sizes, $s$, from
microscopic grains to asteroid-sized bodies, if they are placed within the
host clump.  As we shall see, due to the aerodynamic drag that these solids
suffer, small grains ($s \simlt 0.1 - 1$ cm) are nearly frozen in with the
gas, so have to follow its motion, whereas bodies larger than $s \sim 1$ km
experience negligible friction with the gas. Bodies in the intermediate size
range are most likely to end up joining the solid core. The smaller and the
larger objects may survive the gas clump disruption and become either planet
satellites or independent bodies in heliocentric orbits.

To quantify this discussion, we model the host gas clump in this section only
as in the analytical model of \cite{Nayakshin10c}, which shows that the outer
radius of the host clump is approximately independent of the clump's mass,
$r_{\rm hc} = 0.8$~AU~$k_*^{1/2} (10^4 \mbox{yr}/t_{\rm hc})^{1/2}$, where
$t_{\rm hc}$ is the host clump's age, and $k_*$ is dimensionless opacity
\citep[cf. \S 2.1 of][]{Nayakshin10c}. The age of the host clump chosen for
the representative calculation below is $10^4$ years, the mass is $10 M_J$,
and opacity $k_* = 1$. Note that in detail the structure of the host clump is
different from that of the gas clumps we consider in the numerical of the
paper, but the main conclusions that we reach in this section are
qualitatively unaffected by this.

For simplicity, clump rotation is neglected in this section, but we note that
results for rotating gas clumps are actually quite similar as long as gas
pressure is the main means of support against gravity for the host clump
\citep[in the opposite case the gas clump would be unstable to various fluid
  instabilities; see, e.g., chapter 7 of][]{ShapiroTeukolsky83}.

Our goal here is to calculate how long it takes for a grain of size $s$ and a
given initial condition for the grain position and velocity to settle to the
centre of the clump due to aerodynamic friction with the gas. To achieve this
goal, we solve for the grain's motion within the host clump using the formulae
of \cite{Weiden77} for the aerodynamic forces acting on the grains. This
allows us to calculate the radial sedimentation velocity of the grain, $v_r$,
and define the sedimentation time as
\begin{equation}
t_{\rm sed} = {r \over |v_r|}\;.
\label{tsed}
\end{equation}
where $r$ is distance to the clump's centre.  For solids released from rest
results of such calculations are presented in fig. 2 of
\cite{NayakshinEtal11a}. Here we present a very similar calculation for dust
grains that are on initially circular orbits around the host gas clump's
centre.

\begin{figure}
\centerline{
\psfig{file=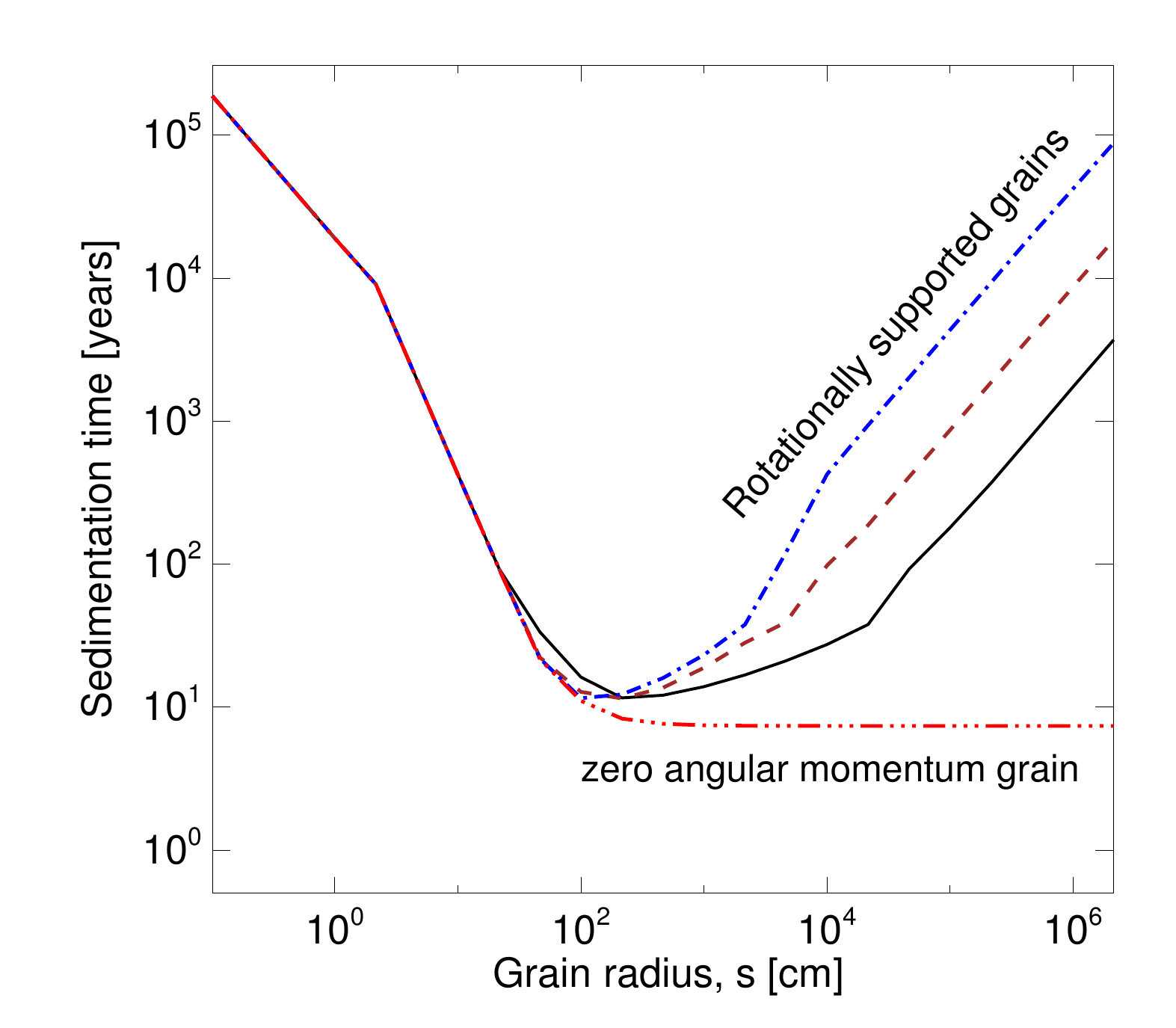,width=0.49\textwidth,angle=0}}
\caption{Sedimentation times versus grain size for grains set initially on
  circular orbits, except for the red dash-triple-dot curve which is for a
  grain released from rest.}
\label{fig:tsed}
\end{figure}

Figure \ref{fig:tsed} shows the sedimentation time scale for grains released
on circular orbits with initial radii $r$ of 0.02, 0.1 and 0.5 times $r_{\rm
  hc}$ for the blue dot-dashed, brown dashed and black solid curves,
respectively. For comparison, the red dash-triple-dot curve shows the grain
sedimentation time for a grain released from rest, as in
\cite{NayakshinEtal11a}.

Without rotation, grains larger than a few cm manage to sediment to the centre
of the gas clump within its age. Smaller grains are tightly bound to the gas,
and if the host is disrupted, these grains are released into the disc
around the protoplanet. The grains closest to the solid core could have been
thermally reprocessed into crystalline materials; mixing these with surrounding
disc's ices and incorporating into comets may explain their puzzling
compositions \citep{Vorobyov11a,NayakshinEtal11a}. 

In a purely spherical geometry with no rotational support for grains, large
grains ($s\simgt 1$ m) fall into the centre and join the solid core on the
free-fall time, which for a constant density model clump is constant with
radius (cf. the horizontal part of the red dash-triple-dot curve).

For grains on circular orbits, however, centrifugal force balances gravity,
and the grains sediment only because of a gradual angular momentum loss due to
aerodynamic forces. These forces become progressively less important for
larger grains, and therefore the grain sedimentation time increases with $s$
again (cf. the black, the brown and the blue curves in
Fig. \ref{fig:tsed}). As a result, grains larger than 1-10 km may orbit the
centre of the clump for times comparable with the clump's lifetime (which we
assume of the same order as the clump's age here, as presumably the clump
continues to migrate inward at roughly same migration speed).

Concluding, we see that solid objects larger than a few km in size do not
necessarily contribute to growth of the solid planetary core as it may take
too long for these bodies to spiral into the core: the host clump is likely to
be disrupted before such an inspiral occurs.

\subsection{On planetesimal birth and sizes}\label{sec:birth}

In this paper we do not simulate the phase of planetesimal formation due to
numerical challenges. Previous simulations of the whole proto-planetary disc
by \cite{ChaNayakshin11a} lacked numerical resolution (to resolve smaller
objects) and resulted in formation of single self-bound massive ($M \simgt$ a
few $\mearth$) clusters of dust particles inside the gas clumps. These were
held from a further self-gravitational collapse artificially by employing a
finite gravitational softening length in the simulations. Increasing
numerical resolution further (that is, the number of SPH and dust particles)
by up to several orders of magnitude is needed to resolve small, e.g., $\simlt
0.01$ AU scales, on which planetesimals we are interested here could
form. This is beyond our present numerical capabilities.

Figure \ref{fig:tsed} shows a difficulty for formation of planetesimals in the
TD scheme, very much similar in nature to the well known ``metre-size
barrier'' for the planetesimal growth in the protoplanetary discs
\citep{Weiden77}. In the CA theory, aerodynamic coupling with the gas makes
$\sim$~metre-sized boulders migrate radially inward in the disc, so that they
are probably lost to the protostar before they become planetesimals.

Similarly, Figure \ref{fig:tsed} shows that solids of intermediate sizes
migrate radially inward very rapidly even if they are initially set on
circular orbits around the centre of the host gas clump. Such objects thus
must join the massive solid core there.  To arrive at km-sized and larger
bodies that suffer much weaker aerodynamic drag, smaller grains must thus
somehow ``jump'' from being small to being large without suffering the
aerodynamic drag.

We leave a detailed study of this issue to a future paper, but make some
suggestions on how this difficulty may be avoided. First of all, again
reminiscent of the well known ideas in the CA theory context, channels for a
rapid growth of solids may be available: via self-gravitational instability
\citep{Safronov72,GoldreichWard73} of dust-dominated regions (although not in
necessarily in the shape of a disc), or via turbulence
\citep[e.g.,][]{YoudinGoodman05,JohansenEtal07,CuzziEtal08}.

Secondly, and unlike the CA ``metre-size barrier'' problem, here it is much
less clear that aerodynamic drag does make all the medium-sized solids to
accrete onto the central object. In the case of a proto-planetary disc (the CA
model), the protostar contains almost all the mass and is thus the natural
central (nearly) point-mass around which the disc rotates. Grains migrating
inward must therefore end up in the star. In the TD case, however, the mass of
the ``massive'' solid core we are thinking about is measured in Earth masses,
which is $\sim 10^{-3}$ of the total host gas clump mass. If there are
turbulent or convective motions in the inner region of the clump (which is
almost a certainty), then the solid core itself will be affected by the gas
motions through its coupling to the gas via gravity of the latter. Thus it may
not ``sit'' in the centre of the clump accreting all the medium-sized solids
sedimenting there. Furthermore, motion of the grains would also be affected,
with turbulent and convective motions of gas dragging the grains around. There
is clearly a limit here; too much of turbulence would prevent grains from
sedimenting into the centre in the first place, but moderate turbulent motions
may be expected to delay grain accretion onto the solid core and lead to
formation of new condensation centres which perhaps would lead to formation of
the larger planetesimals that we study in the rest of the paper.

It would also be desirable to understand just how large the planetesimals
formed inside the host gas clump are likely to be. Following \S 3.6.2 and
3.6.3 of \cite{Nayakshin10a}, we find that the linear extent of the region of
the gas cloud within which the first gravitational collapse of grain-dominated
regions could occur is about $\sim 0.1$ times the size of the gas clump. For
the total grain mass in the collapsing region of $10\mearth$, the expected
velocity dispersion of fragments formed by collapse is then $\sim 0.5$ km
s$^{_1}$.

Now, collisions of large planetesimals of equal size, $a$, would lead to their
fragmentation if the collision velocity ($\sim$ the velocity dispersion
calculated above) is larger than the escape velocity from the surface of the
planetesimal \citep[e.g.,][]{StewartL09}. This would then suggest that
collisions of equal size bodies inside the ``grain sphere'' would split
planetesimals smaller than about 1000 km, while larger bodies would
``stick''. One however need to estimate the frequency of such collisions.
Assuming that the grain sphere fragments into bodies of approximately equal
size, we find that all the planetesimals would have had at least one
catastrophic (shattering) collision within a thousand years inside the grain
sphere for $a\simlt 100$ km. Collisions with smaller bodies -- fragments of
the ``original'' planetesimals -- increase this collision rate estimate
further. We thus conclude that expected sizes of planetesimals surviving the
dense environment of the inner region of the clump are at least a few hundred
km. Presumably both smaller and larger solid objects could then be obtained
over much longer time scales by collisional evolution of the disrupted
population.

\subsection{Planet's influence radius: bound and unbound debris}\label{sec:influence}

Having understood the range of body sizes ($s\simgt 1$km) that may be
considered as independent for the duration of the host clump, we now shift out
attention to what may happen when the host clump is disrupted. In detail the
answer on this question is quite complicated as it depends on the nature of
planetesimal orbits within the clump and the exact way the clump is
disrupted. However, for not too eccentric orbits, we can divide the
planetesimal population on those that are ``close'' and those that are ``far''
from the solid planetary core. The former have a fair chance of remaining
bound to the solid core when the host clump is disrupted, and the latter
become unbound independent objects.

To quantify this, consider the simplest setting, placing a solid core of mass
$M_p$ in the centre of the gas clump and setting planetesimals on circular
orbits around the planet. The rotation of the planetesimal's disc is assumed
prograde with respect to the orbit of the host clump around the star. The
circular speed of the planetesimals around the planet is given by
\begin{equation}
v_{\rm circ}^2 = G {M_p + M(r)\over r}\;,
\label{vcirc}
\end{equation}
where $M(r)$ is the gas mass enclosed inside radius $r$ (distance from the
centre of the host clump). 

If planet's mass is low enough to build up a massive self-gravitating
atmosphere around it \citep[which is expected to be the case for an adiabatic
  young gas clump where the critical core mass is above 100
  $\mearth$; see][]{PerriCameron74}, then the gas density is nearly constant in the
clump's centre. Let us call that central density $\rho_0$; thus $M(r)\approx
(4\pi /3) \rho_0 r^3$. We can now define the planet influence radius, $r_i$,
such that $M(r_i) = M_p$. Apparently,
\begin{equation}
r_i = \left[{3 M_p\over 4\pi \rho_0}\right]^{1/3}\;.
\label{ri}
\end{equation}
Note that since $r_{\rm hc} = (3 M_{\rm hc}/4\pi \rho_{\rm mean})^{1/3}$,
where $\rho_{\rm mean}$ is the mean density of the host clump, then we have
\begin{equation}
r_i \approx r_{\rm hc} \left({M_p \over M_{hc}}\right)^{1/3} \left({\rho_{\rm
    mean}\over \rho_0}\right)^{1/3} = 0.18\; r_{\rm hc} 
\left({\rho_{\rm mean}\over \rho_0}\right)^{1/3}
\label{ri_app}
\end{equation}
Solids inside $r_i$ are gravitationally bound mainly to the planet, whereas
those outside are bound by the enclosed gas. We expect that when the host
clump is disrupted, objects within $r_i$ remain bound to the solid planet
whereas objects outside this radius become unbound from the planet.

\subsection{Post-disruption orbits for unbound debris}\label{sec:theory}

We now build an approximate analytical theory for the orbits of the
planetesimals after the host clump tidal disruption. To this end we assume
that gas clump disruption is instantaneous. We also assume that solids within
$r_i$ remain bound to the planet and continue to follow its motion around the
central star, and we concentrate here only on solids outside the influence
radius.

Before the disruption, the position of a planetesimal with respect to the star
is given by $\mathbf R+\mathbf r$, where $\mathbf R$ is the vector connecting
the star with the centre of the gas clump (solid planet's location), and
$\mathbf r$ is the vector connecting the centre of the host clump with the
given planetesimal. The instantaneous velocity of the planetesimal is also
comprised of two parts, one due to the heliocentric Keplerian motion of the
gas clump, $\mathbf V$, and the other given by the circular prograde rotation
velocity, $v_{\rm circ}$, around the centre of the host clump.

We now call $\mathbf{V + \Delta v}$ the planetesimal's velocity right after
the host clump disruption. Physically, $\mathbf{\Delta v}$ is the velocity
with which planetesimal becomes unbound, and may be called a velocity
``kick''. We expect that $0 < \Delta v < v_{\rm circ}$ for the energy
conservation reasons. 

We shall now calculate the new heliocentric orbit of the planetesimal by
considering its specific energy and angular momentum.  The specific energy and
specific angular momentum of the host gas clump before the disruption are $E_0 =
-GM_*/2R$ and $\mathbf{L_0 = R\times V}$, respectively.  The post-disruption
specific energy, $E$, and specific angular momentum, $ \mathbf L$, of the
planetesimal are
\begin{eqnarray}
E &=& -\frac{GM_*}{|\mathbf R+\mathbf r|} +
\frac{1}{2}\left|\mathbf V + \mathbf{\Delta v}\right|^2,\\
\mathbf L &=& (\mathbf R+\mathbf r) \times (\mathbf V + \mathbf{\Delta v}),
\end{eqnarray}
where, $M_*$ is the mass of the central protostar. Decomposing these
expressions into Taylor series with respect to small parameters $r/R$ and
$\Delta v/V$, up to the linear terms only, and writing $E \approx E_0 + \delta
E$, $L^2 \approx (L_0 + \delta L)^2$ (the latter is possible since
$\mathbf{\Delta v}$ is in the clump's orbital plane by setup), we have
\begin{equation}
\delta E = V^2 \left({\mathbf{R \cdot r} \over R^2} + {\mathbf{V \cdot \Delta
    v} \over V^2}\right)\;. 
\label{deltae_app}
\end{equation}
From this we see that the semi-major axis of the planetesimal's orbit after
disruption is 
\begin{equation}
a = \frac{GM_*}{2|E|}\; = R \left(1 + {\delta E\over |E_0|}\right) = R
\left[ 1 + 2 \left({\mathbf{R \cdot r} \over R^2} + {\mathbf{V \cdot \Delta v}
    \over V^2}\right)\right]\;.
\label{a_linear}
\end{equation}
Similarly, we can write
\begin{equation}
\mathbf{\delta L \; \approx \;  r \times V + R\times \Delta v}\;.
\label{l_linear}
\label{deltaL1}
\end{equation}
Now, using vector identity 
\begin{equation}
\mathbf{[A\times B]\cdot [C\times D] = (A\cdot C)(B \cdot D) - (A \cdot D) (B
\cdot C)}\;,
\end{equation}
we find that
\begin{equation}
\mathbf{L_0 \delta L =} V^2 \mathbf{R \cdot r} + R^2 \mathbf{V \cdot \Delta v}
\end{equation}
Comparing this expression with equation \ref{deltae_app}, we observe that
\begin{equation}
{\delta E\over E_0} = - 2 {\mathbf{\delta L\cdot L_0}\over L_0^2} = -2 {\delta
  L \over L_0}\;,
\label{de_dl}
\end{equation}
where we also recalled that the orbital plane of the planetesimals does not
change since the proto-disc of planetesimals rotates in the prograde direction
in the orbital plane of the host clump by assumption.

The  eccentricity of the planetesimal's orbit is
\begin{equation}
e^2 = 1+\frac{2E L^2}{G^2M_*^2}\;,
\label{eq:ana_e}
\end{equation}
which can be decomposed to become
\begin{equation}
e^2 \approx - {\delta E\over E_0} - 2 {\delta
  L \over L_0} - 2 {\delta E\over E_0} {\delta
  L \over L_0} - \left({\delta
  L \over L_0}\right)^2\;.
\end{equation}
Using equation \ref{de_dl}, we find that the linear terms cancel, and the
always positive result is
\begin{equation}
e^2 = {3 \over 4} \left({\delta E \over E_0}\right)^2\;.
\label{ecc1}
\end{equation}

\subsubsection{Eccentricity -- semi-major axis correlation}\label{Sec:e_a}

We can eliminate $\delta E$ and $E_0$ in favour of orbital elements. Since $E_0 =
-GM_*/2R$, and $a = GM_*/2|E|$, equation \ref{ecc1} actually shows that
\begin{equation}
e\approx \left({3 \over 4}\right)^{1/2} {|\delta a| \over a_0}\;,
\label{ecc_a}
\end{equation}
where $a_0 = R$, and $\delta a \equiv a - R$ is the difference in the
semi-major axis of the planetesimal's post-disruption orbit and that of the
host gas clump before disruption.

Equation \ref{ecc_a} predicts a correlation in the orbital elements of
planetesimals after disruption: the belt of planetesimal debris left after the
host gas clump disruption should have circular orbits at the centre, $a=a_0$,
and increasingly eccentric orbits towards the belt's edges.

We note one rather interesting feature of equation \ref{ecc_a}: it does not
depend on the properties of the host gas clump {\em or} the solid
protoplanetary core inside. The eccentricity -- semi-major axis correlation
should thus be a general property of the debris rings left after disruption of
the host gas clumps (as long as the pre-disruption orbit of the clump is
  nearly circular).

We also note that for the case of the proto-disc of planetesimals {\em not
  coinciding} with the orbital plane, the inclination of the planetesimal
orbits, $i$, after the disruption is non zero but can be shown to be small.

\subsubsection{Rings versus discs}\label{Sec:rings}

Despite the universal correlation we just discussed, the properties of the
host gas clump are still imprinted on the {\em range} of possible planetesimal
orbits through the fact that the pre-disruption protoplanet is finite in
spatial extent, and that the maximum velocity kick to a planetesimal released
by the clump disruption, $\Delta v$, is also finite. Referring to equation
\ref{a_linear}, we see that the width of the ``disruption ring'' -- the ring
occupied by the planetesimals left over after the clump disruption -- is
\begin{equation}
w\equiv {a_{\rm max} - a_{\rm min} \over a_0} = 4  \max \left| {\mathbf{R \cdot r} \over
  R^2} + {\mathbf{V \cdot \Delta v} \over V^2} \right|
\label{a_width}
\end{equation}
Noting that $\Delta v \sim v_{\rm circ} \propto r$ within a constant density
host clump, and that the maximum possible value of $r$ is the Hill's radius of
the host clump, $r_h$, we can summarise this prediction by writing
\begin{equation}
w  \approx 4 \zeta {r_h \over a_0}\;,
\label{zeta}
\end{equation}
where $\zeta$ is a dimensionless number probably smaller than but comparable
to unity. For the setup of this paper, in particular, where $r_h = 4.7$ AU and
$a_0 = 40$ AU, we have
\begin{equation}
w \approx 0.5 \zeta\;.
\label{w_predict}
\end{equation}
We shall see below that $\zeta \approx 0.7-1$ for the three simulations
performed in the paper. In general this parameter will depend on how
concentrated the distribution of planetesimals is within the host protoplanet
before the disruption, with more compact distributions leading to smaller
$\zeta$. It would also vary, likely increase, if planetesimal orbits in the
pre-disruption clump are eccentric.

\section{Numerics}\label{sec:numerics}

\subsection{Method} \label{sec:method}

We now turn to hydro/N-body simulations for a more detailed investigation of
the problem. SPH \citep{Gingold77,Lucy77} is a Lagrangian simulation algorithm
well suited for irregular and self-gravitating systems. SPH has been applied
to a variety of astrophysical contexts \citep{Monaghan92,Springel10}.  In this
paper, we use \textsc{gadget-3}, an updated version of the SPH/N-body code
presented by \cite{Springel05}.

In both of the disruption scenarios that we study in this paper, host clump
disruption occurs on a time scale shorter than the Kelvin-Helmholtz time of the
clump. An adiabatic equation of state for the gas is thus sufficient for our
purposes. As hydrogen is molecular for temperatures smaller than $\sim 2000$
K, we chose the polytropic index, $n$, to be $n=5/2$, which corresponds to the
ratio of specific heats of $\gamma=1.4$. The number of SPH particles in each
of the simulations presented is $N_{sph} = 10^{6}$, so that the mass of each
particle is $5\times 10^{-6} M_J$.

The protostar, the planet and the planetesimals are all modelled as N-body
particles of appropriate masses. Accretion onto the planet is not allowed, but
accretion onto the protostar is simulated with the sink particle approach
\citep[as in][]{ChaNayakshin11a}. This is necessary to prevent the build up of
very short time step SPH particles near the star due to a non-negligible
artificial viscosity of the code in this low density region, which is
additionally of little interest to our study.

The N-body particles interact with gas only through gravity (except for gas
accretion as described above). We set the total mass of the planetesimal disc
to $M_d = 0.1$M$_\oplus$. This mass is shared equally between $N_{\rm pl} =
2\times10^4$ particles. The mass of each is thus $\sim 3\times 10^{22}$ g, and
the corresponding radius is about 200 km. For such low masses, gravitational
interactions between planetesimal particles and their effect onto the planet
and the gas are negligible, even though these interactions are included in the
calculation. The large linear size of the planetesimals allows us to neglect
the aerodynamic coupling between them and the gas completely.

\subsection{Initial conditions and disruption method} \label{sec:init_cond}

Below we present three numerical simulations. The first of these, labelled U0,
uses the initial condition for the polytropic gas cloud described in \S
\ref{sec:tidal}. In this case the host gas clump exactly fills its Roche lobe
at the beginning of the simulation. The clump is thus disrupted due to tidal
forces of the star only.

In the other two cases, the host clump radius, $r_{\rm hc} = 1.84$ AU, is
initially smaller than the Hills radius, $r_h=4.7$ AU. The host clump central
temperature is $500$ K for this initial condition. As explained in \S
\ref{sec:explosion}, we assume that at $t=0$ the central solid core releases a
given amount of energy which is then instantaneously added to the internal
energy of the gas particles within radius $r_{\rm ej} = 0.2$ AU (chosen
somewhat arbitrarily). The inner region is then over-pressured compared to its
surroundings, which drives an expansion that puffs up the whole gas clump.

In \S \ref{sec:explosion} we found that injecting energy $\Delta U_0 = 0.123
U_0$, where $U_0$ is the total internal energy of the gas protoplanet, should
increase $r_{\rm hc}$ to $r_h$.  This derivation assumed that the clump keeps
its polytropic structure even after the energy injection, which is clearly not
actually correct for the equation of state we use. We therefore performed two
simulations with energy injection bracketing the critical injection energy
$\Delta U_0$. These are labelled U10 and U15, so that $\Delta U = $ 10\% and
\%15 of $U_0$ for the two simulations, respectively. We find that both of
these simulations resulted in the disruption of the host gas clump, although
the process was much faster in U15 than in U10. There are interesting
differences between the orbital parameters of the planetesimals in U10 and
U15, making both simulations worth presenting here.

The exact outcome of the gas protoplanet disruption depends on a number of
assumptions about the pre-disruption stage, such as (i) the exact location of
the planet at the moment of the gas protoplanet disruption (which needs not be
the centre of the gas clump in general); (ii) the duration of the disruption
process (rapid or slow); (iii) the distributions and orbits of solids within
the clumps, and their total mass, which may start influence the dynamics of
planetesimals if their total mass is comparable or larger than that of the
planet, and (iv) the orbit of the gas clump around the protostar.

The parameter space of the problem is too large to cover in this first
study. Therefore, we only aim here at learning the most roburst features of
the results, which we hope will be generally correct within a factor of a few
despite all the uncertainties pointed out above.

The planetesimal particles are placed in a geometrically thin disc on circular
orbits around the centre of the gas clump with prograde velocity given by
equation \ref{vcirc}.  The outer radius of the planetesimal disc, $r_d$, is
$0.4$ AU for the simulations U10,U15 and $1$ AU for simulation U0,
respectively (to scale approximately as $0.2 r_{\rm hc}$ for all the
cases). The inner radius of the planetesimal disc, $r_{in}$, is set to 10$\%$
of $r_d$. Introduction of the inner disc cutoff at $r_d$ does not change our
results at all but speeds up simulations considerably. Particles at $r \simlt
r_{in}$ are bound very strongly to the solid core (planet), so that disruption
of the host gas clump hardly changes their orbits. These ``uninteresting''
planetesimals have short time steps and are expensive to simulate.

The planetesimal disc has the intitial disc surface density profile of
$\Sigma_{\rm pl} (r) \propto r^{-2}$. As partciles in the disc do not
interact significantly due to their low masses, behaving as test particles,
such a choice for $\Sigma_{\rm pl} (r)$ has no consequences for the results
but allows an approximately uniform parameter sampling in terms of the initial
distance $r$ to the solid planet.

Initially, the gas host clump rotates around the protostar with the Keplerian
velocity, $\sqrt{GM_*/R}$.  The orbital period of the initial orbit, $P_0$, is
$\approx 253$ yrs. We use $P_0$ as a unit of time in presenting the results
below. All of the simulations were run until time $t=10$, i.e., 10 initial
orbits of the clump.

\section{Dynamics during host protoplanet disruption}\label{sec:gas}

\subsection{Simulation U15} \label{sec:u15}

\begin{figure*}
\centerline{
\psfig{file=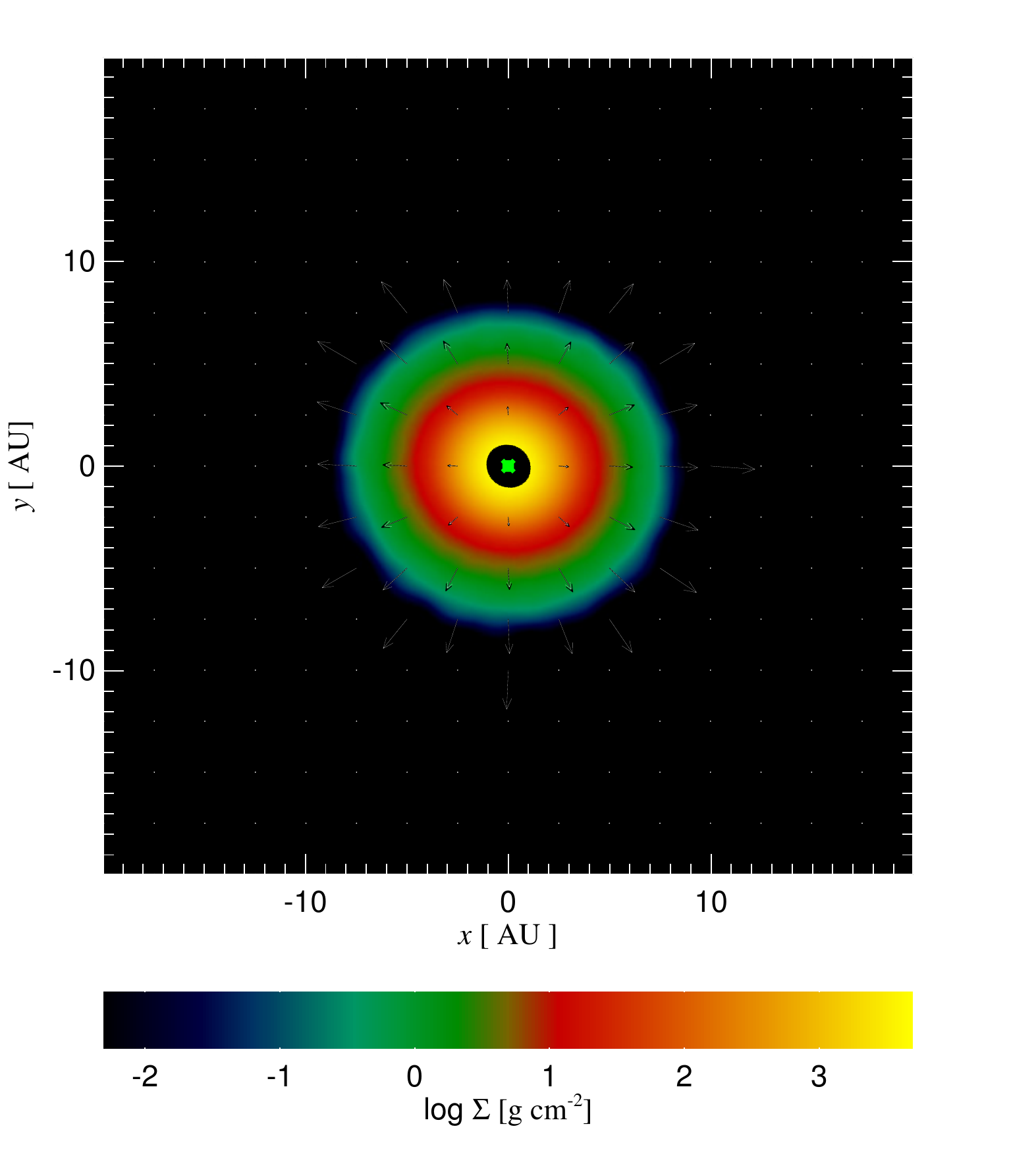,width=0.49\textwidth,angle=0}
\psfig{file=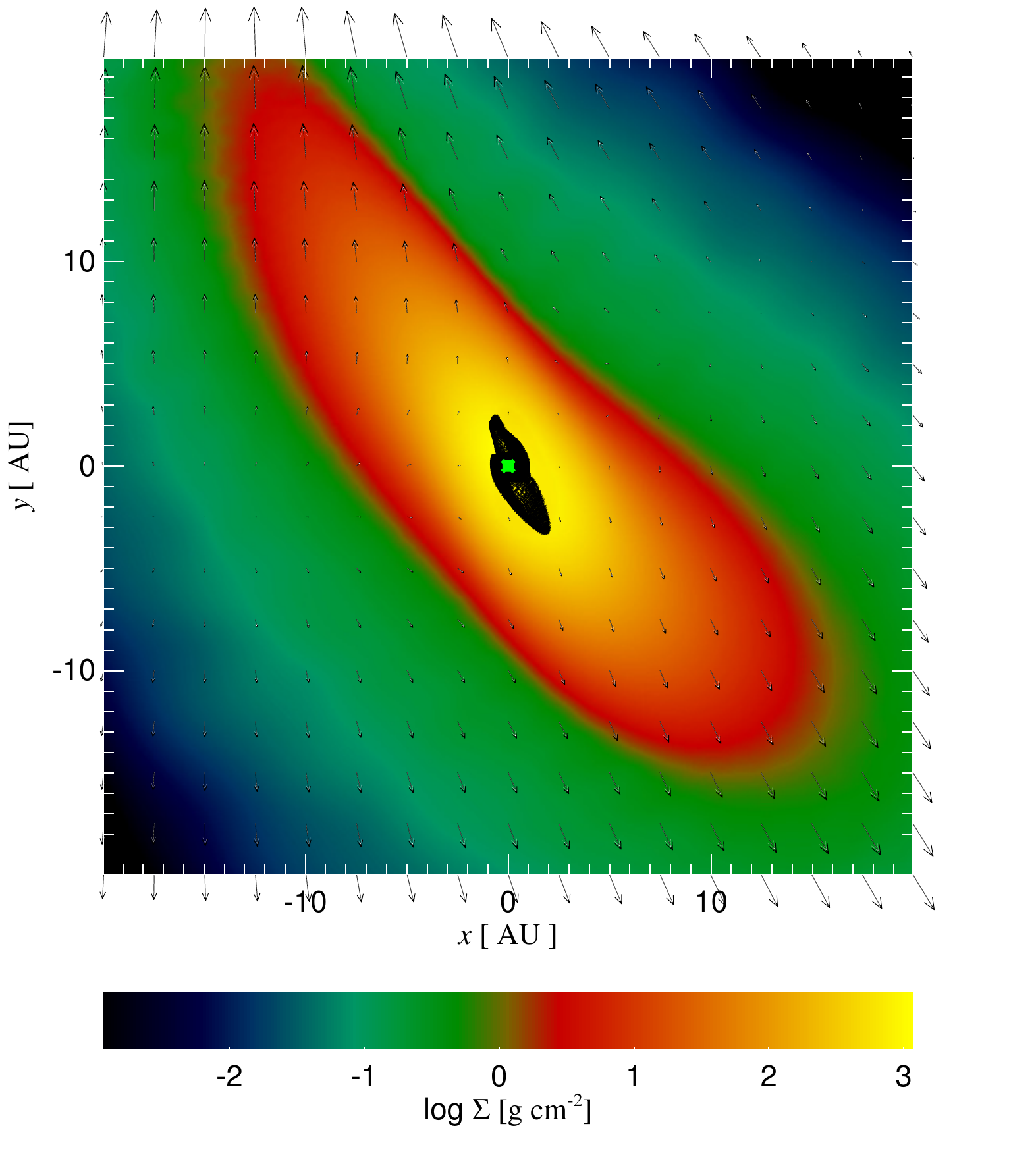,width=0.49\textwidth,angle=0}}
\caption{The top view of the gas surface density map for simulation U15 at
  time t=0.06 (left panel) and $t=0.25$ (right panel). The map is centered on
  the position of the solid planet (core), which is shown with the thick green
  symbol. Vectors show the velocity field, also centered on the velocity of
  the planet. The black dots show locations of individual planetesimals.}
\label{fig:OFF00EB015VY_early}
\end{figure*}

In this simulation, the energy injected into the gas exceeds the critical
energy needed to disrupt the host clump
(cf. eq. \ref{deltau_disr}). Therefore, a rapid disintegration of the gas
clump is expected. Fig. \ref{fig:OFF00EB015VY_early} shows the projected gas
column density at two early stages of the simulation, $t = 0.06$ (left panel)
and $t = 0.25$ (right panel). The projection is done along the $z$-axis, e.g.,
along the direction normal to the host clump's orbital plane. The figures are
centered on the position of the solid core (planet), shown with the thick
green symbol at the centre of each panel. Vectors show the gas velocity field,
likelywise centered on the solid planet's velocity.  The black dots are
individual planetesimals.

Initially (left panel), clump expansion appears nearly spherically symmetric
even though the outer gas shells did overflow the Roche lobe by that
time. However, the flow beyond $r_h$ is controlled more and more by the star's
tidal field. This is evident in the right panel, where the flow becomes highly
assymetric at large $r$. The star is positioned exactly North of the clump in
the right panel.  One observes a flow of gas towards the star at the upper
left corner of the panel, and away from the star in the lower part of the
panel.  The planetesimal disc is also becoming affected, visually following
the motion pattern of the gas. 

The host clump disruption is indeed dynamic as the right panel corresponds to
time just slightly later than one dynamical time, defined as $P_0/2\pi$.

\begin{figure*}
\centerline{
\psfig{file=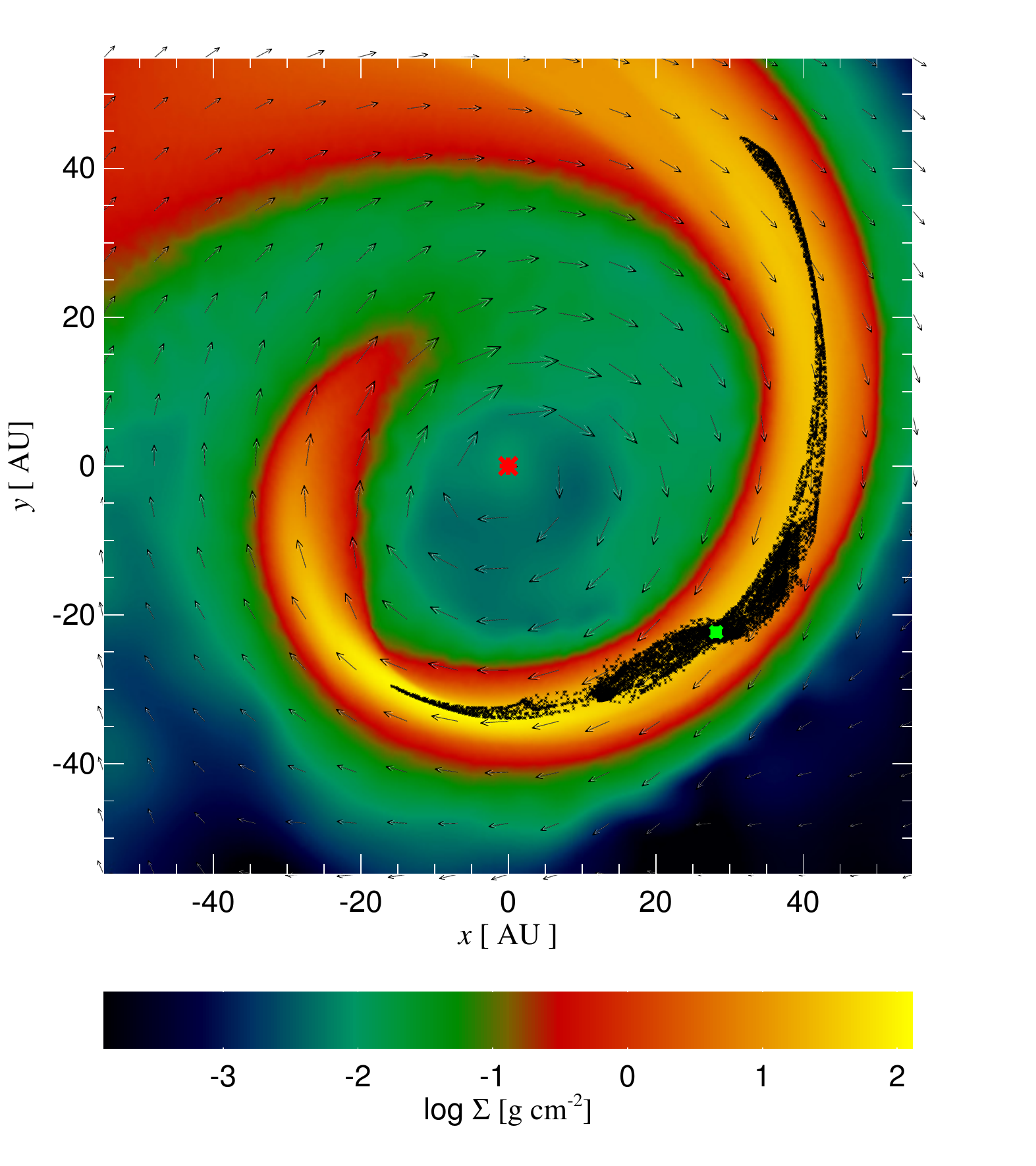,width=0.49\textwidth,angle=0}
\psfig{file=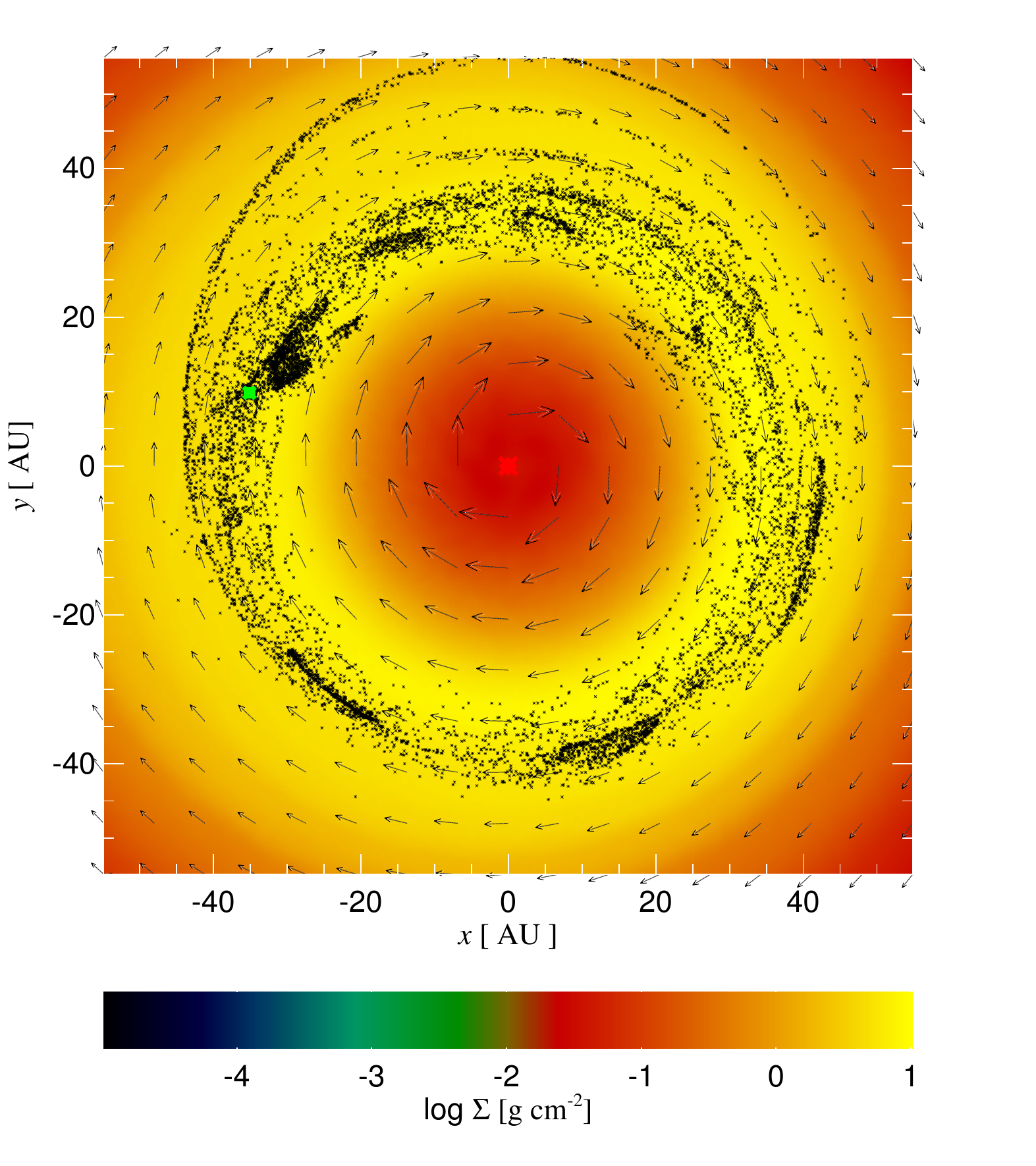,width=0.49\textwidth,angle=0}}
\caption{Same as Fig. \ref{fig:OFF00EB015VY_early} but at later time; $t=1$
  (left panel), and $t=10$ (right panel), except that the center of the
  coordinate system is now the protostar (the red symbol at the center of the
  panels). Note that planetesimals initially continue to follow the orbital
  motion of the densest parts of the gas streams at $t=1$ but separate out at
  $t=10$ when the gas streams are dispersed and circularised into a ring.}
\label{fig:OFF00EB015VY_late}
\end{figure*}

Fig. \ref{fig:OFF00EB015VY_late} shows the same as
Fig. \ref{fig:OFF00EB015VY_early}, but now at times $t=1$ and $10$ in the left
and the right panels, respectively. The figure presents a larger field of view
centered onto the star now. The velocity field is also centered onto the star
rather than the planet. Note in the left panel that the planetesimals
disrupted off the initial gas clump continue to follow orbits similar to that
of the densest regions of gas. This is hardly surprising given that potential
due to the host clump initially dominates the planetesimals' orbits (until the
host clump is completely disrupted). At late times (see the right panel), the
host clump is disrupted into a ring centered about the initial clump's
separation from the star. The planetesimals are also spread into a ring-like
feature which shows  several streams. The streams appear to have some eccentricity to
them. We also note that the fine structure of the streams is now not
correlated to the gas component. This is probably caused by the gas
gravitational potential being smoothed out due to circularisation of the gas
flow, leading to a reduced gas gravitational force on the planetesimals.

\subsection{Simulation U10}\label{sec:u10_gas}

As derived in \S \ref{sec:explosion}, the critical injection energy sufficient
to inflate the clump to the point of its tidal disruption is $\Delta U_0 =
0.123 U_0$ for the parameters of the clumps we consider. Therefore, if the
approximate theory of \S \ref{sec:explosion} were correct, the clump would not
be disrupted at all in this particular simulation as $\Delta U = 0.1 U_0 <
\Delta U_0$. However, the argument given by equation \ref{deltau_disr} assumes
that the energy injected into the central regions of the gas is shared by the
whole clump, so that gas in the clump effectively finds itself on a new
(single) polytropic relation. In the simulations, however, the central regions
are shifted to a different polytropic relation by the energy injection,
whereas the rest of the clump remains at the initial one. One may thus expect
the simulation results to differ from the simple analytical prediction.

\begin{figure}
\centerline{\psfig{file=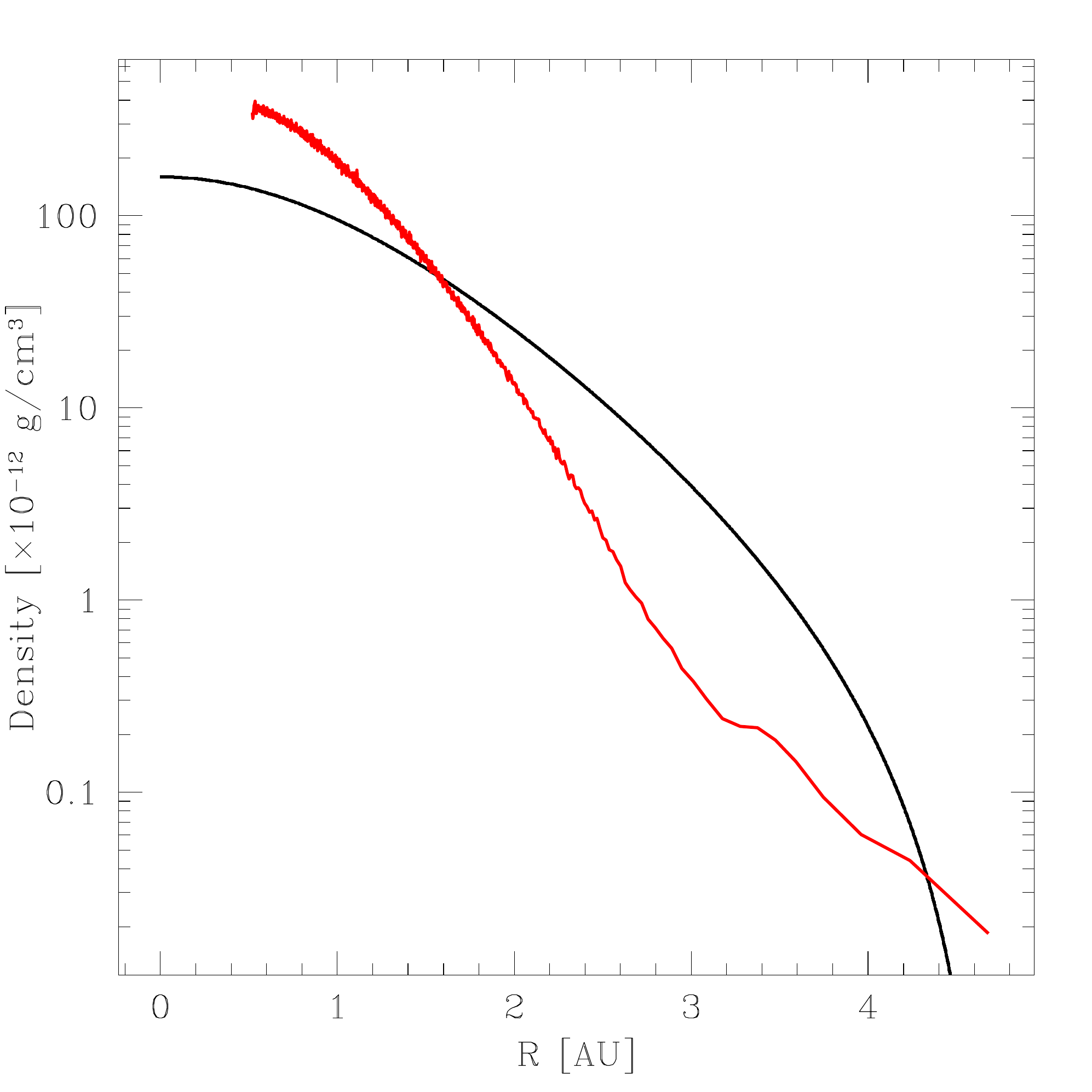,width=0.49\textwidth,angle=0}}
\caption{The internal density structure of the expanding protoplanetary clump
  for simulation U10 ($\Delta U/U_0=0.1$) at time $t=0.05$. The red solid line
  is the actual gas density from the simulation, while the black solid line is
  a polytrope of 5M$_J$ mass and $4.7$AU radius. The latter curve is expected
  based on the approximate theory shown in \S \ref{sec:explosion}. The figure
  shows that a single polytropic relation approximation breaks down after the
  energy injection, although the approximation is still useful in roughly
  determining the critical injection energy $\Delta U_0$.}
\label{fig:OFF00EB010VY_early_internal}
\end{figure}

Figure \ref{fig:OFF00EB010VY_early_internal} shows the radial density
profile of the host clump (red curve) in simulation U10 at time $t=0.05$
and the theoretical polytropic density profile (black curve) with $n=5/2$
inflated by the energy input from the solid core as assumed in \S
\ref{sec:explosion}. One notices significant differences between the two
curves, with a tail of gas material extending further out in the red curve. It
is this tail that allows some material to siphon out of the Roche lobe and
for the gas clump's radius to continue swelling (cf. equation
\ref{mass_radius} on this point) as mass is lost.

\begin{figure*}
\centerline{
\psfig{file=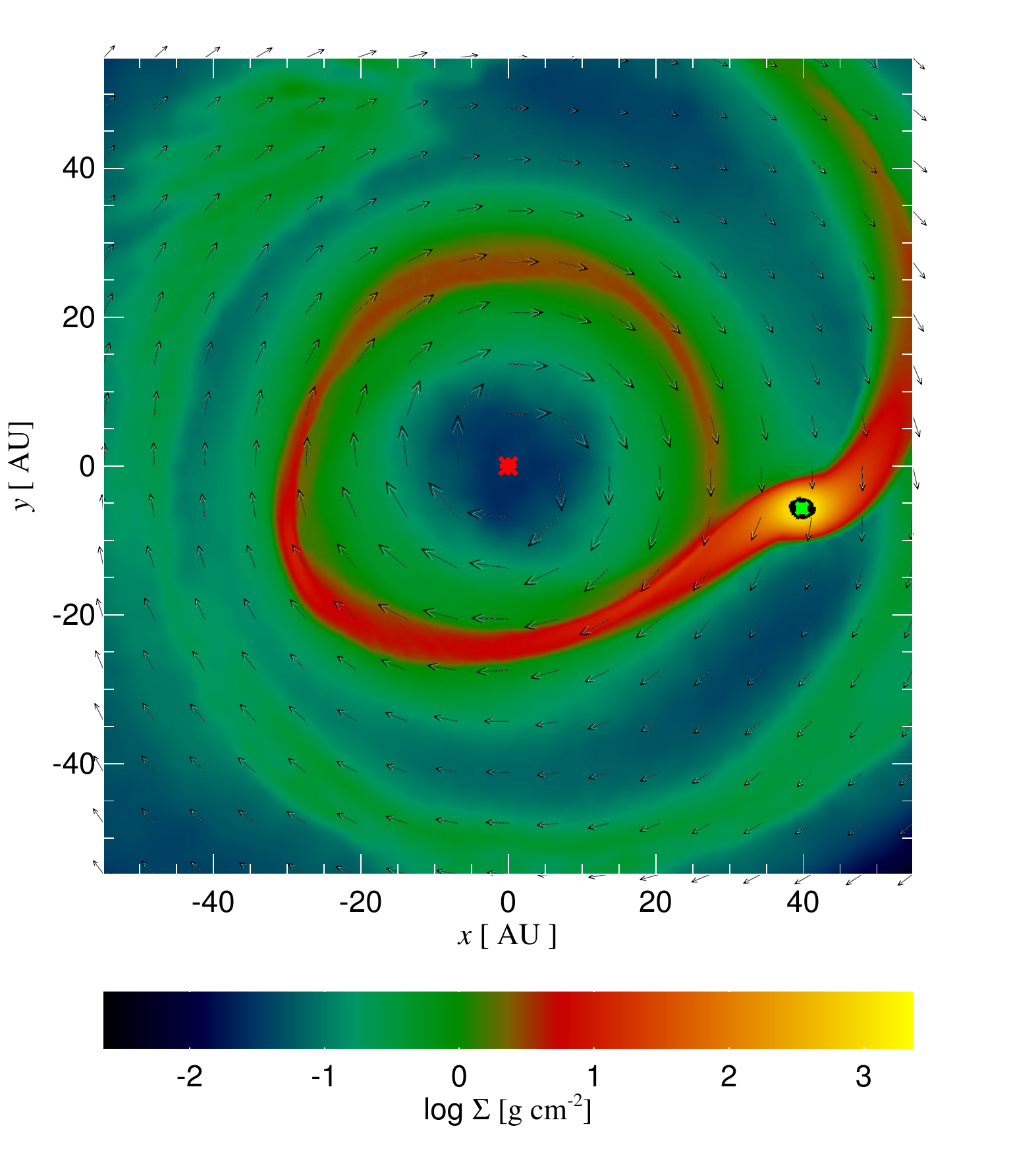,width=0.49\textwidth,angle=0}
\psfig{file=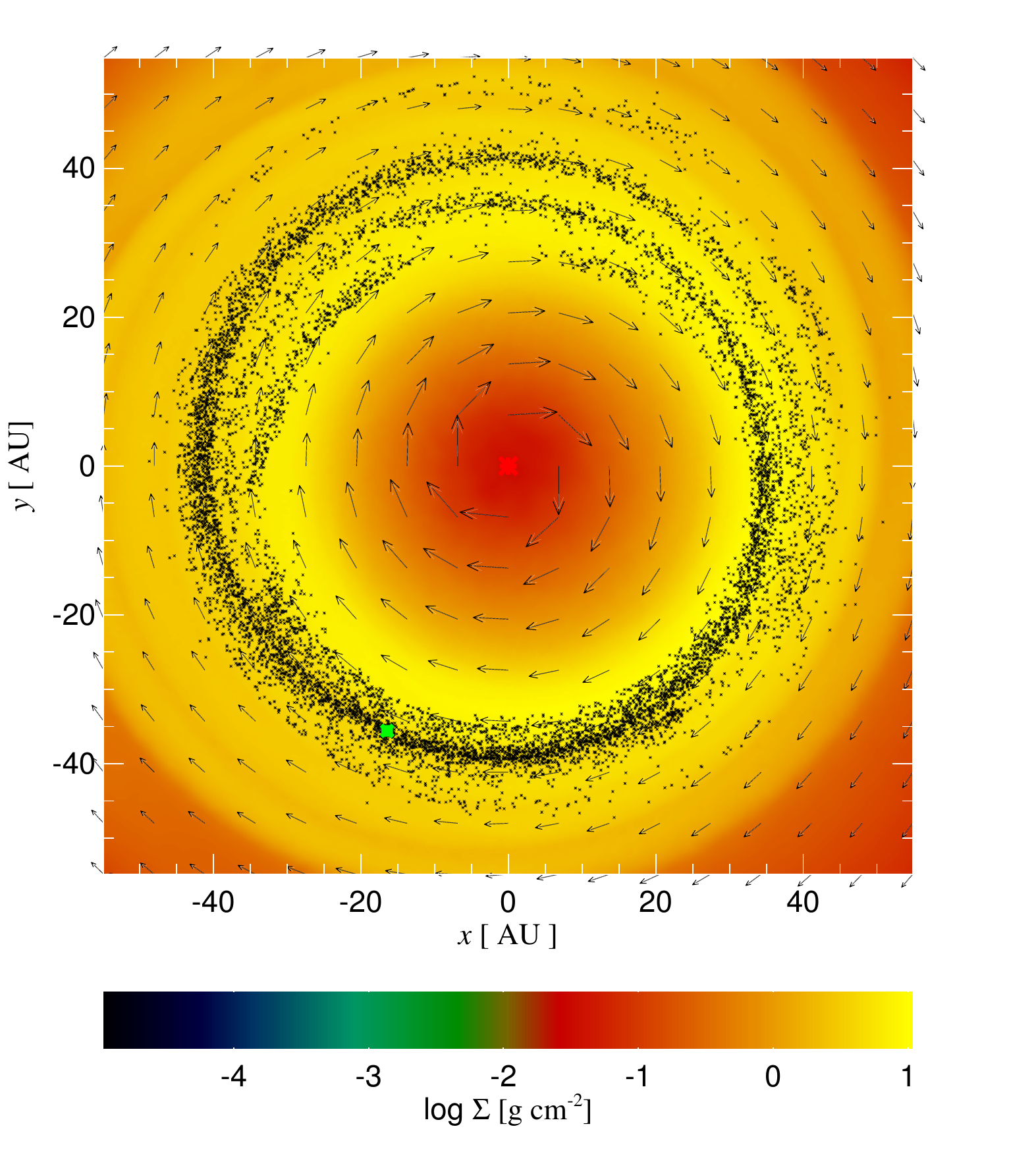,width=0.49\textwidth,angle=0}}
\caption{Gas column density projections similar to Fig.
  \ref{fig:OFF00EB010VY_early_internal} but for run U10 at time $t=3$ (left
  panel) and $t=10$ (right panel). As the planet is less extended than in test
  U15, the disruption process is much more gradual. At $t=3$ (3 whole orbits
  of the host clump around the star), only a small fraction of gas mass is
  lost through the Roche lobe overflow. No planetesimals have yet been unbound
  from the clump. However, at late time (the right panel), the structure of
  the disrupted gas and planetesimal population is quite similar to that in
  U15.}
\label{fig:OFF00EB010VY_3P_snap}
\end{figure*}

The left panel of Figure \ref{fig:OFF00EB010VY_3P_snap} illustrates this
initially slow expansion. The host clump is largely intact at $t=3$ in
U10. This is in stark contrast to simulation U15 in which the host gas clump
was completely obliterated by $t=1$ already (left panel of
Fig. \ref{fig:OFF00EB015VY_late}). However, the host gas clump is eventually
destroyed in simulation U10 also. The gas column density and the distribution
of planetesimals appear to be quite similar at late times in U10 and U15
($t=10$, right panels of Figs. \ref{fig:OFF00EB010VY_3P_snap} and
\ref{fig:OFF00EB015VY_late}, respectively). We shall make a more detailed
analysis of planetesimals' orbits in \S \ref{sec:unbound} below.

\subsection{Simualtion U0}\label{sec:u0_gas}

\begin{figure*}
\centerline{
\psfig{file=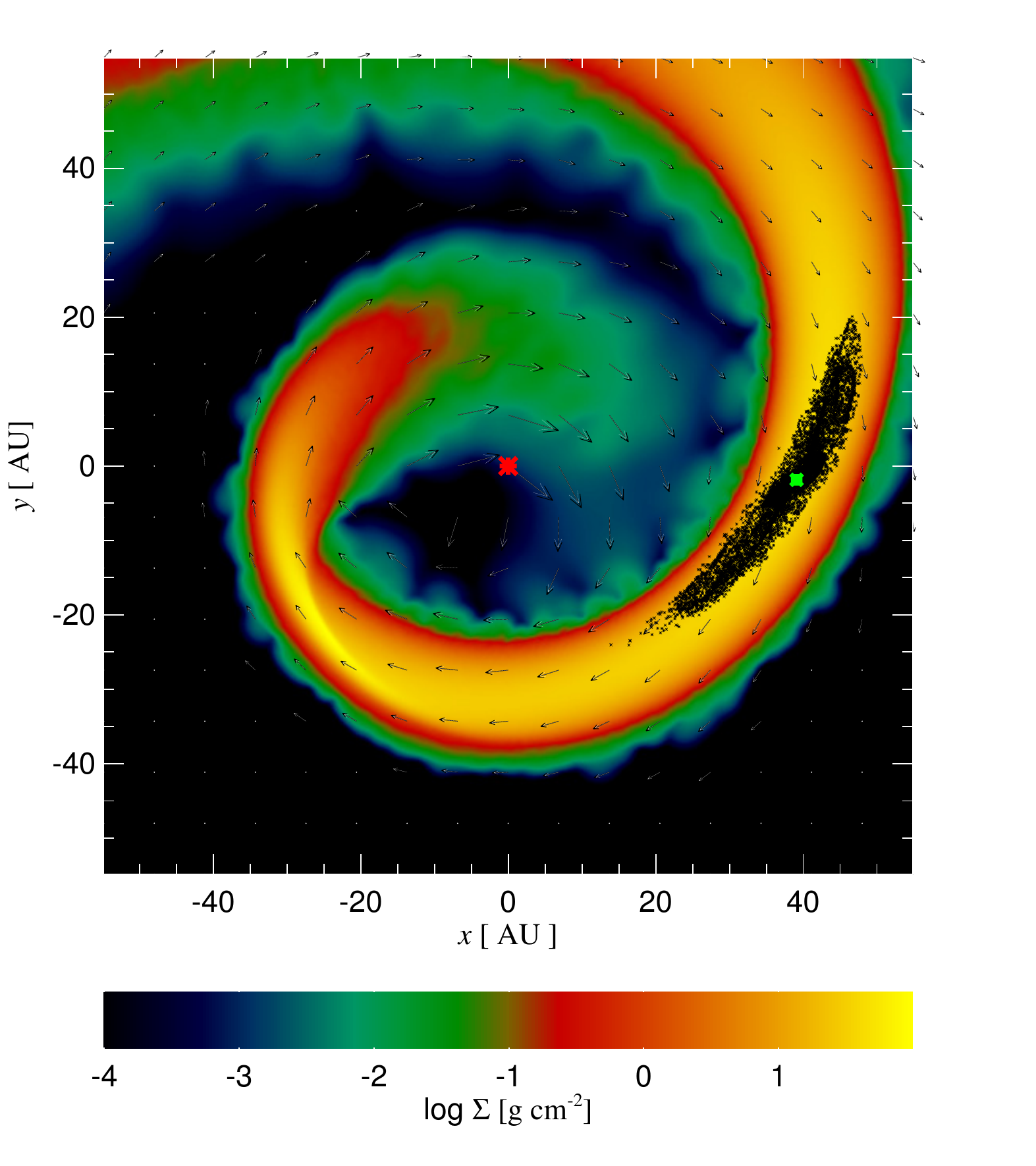,width=0.49\textwidth,angle=0}
\psfig{file=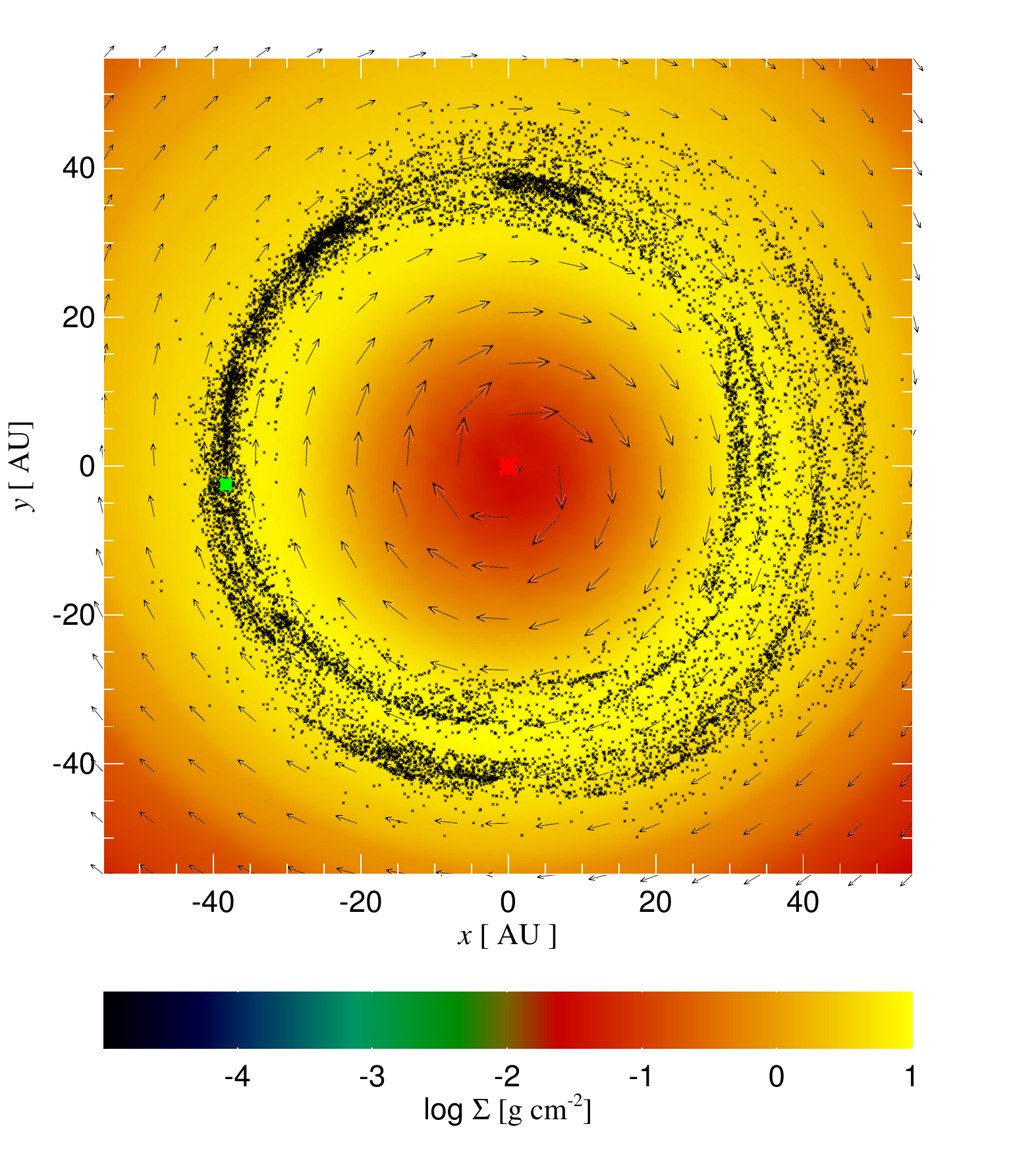,width=0.49\textwidth,angle=0}}
\caption{Same as Fig. \ref{fig:OFF00EB010VY_3P_snap} but for simulation U0,
  the case where the initial cloud is cooler and already fills its Roche
  lobe. The corresponding times are $t=4$ (left panel) and $t=10$ (right
  panel).}
\label{fig:BIG_snap}
\end{figure*}

In this simulation, the gas protoplanet is more extended initially, so that
$r_{\rm hc} = r_h$ at $t=0$. No energy input from the solid core is
assumed. Figure \ref{fig:BIG_snap} shows two snapshots for simulation U0 at
times $t=4$ (left panel) and $t=10$ (right panel) in the same format as
Fig. \ref{fig:OFF00EB010VY_3P_snap}. The disruption process is not as rapid as
in simulation U15 but is a little faster than in U10. The morphology of the
gas flow is different: the disrupted gas spiral at $t=4$ is wider in U0 than
it is in Fig. \ref{fig:OFF00EB010VY_3P_snap}. 

Since the initial disc or planetesimals inside the gas protoplanet is large to
begin with ($r_d = 1$ AU for U0) than for simulations U10 and U15 ($r_d = 0.4$
AU in both), the flow of disrupted planetesimals is initially wider in U0 (see
the banana-shaped feature in the left panel of Fig. \ref{fig:BIG_snap}.) The
end result is at least visually not too dissimilar from runs U10 and U15,
however.

\section{Orbits of the unbound planetesimals}\label{sec:unbound}

We now turn to analysis of the unbound population of planetesimals in terms of
their orbital elements at the end of the simulations. To avoid confusion, tt
must be stressed that, all of our planetesimals remain bound to the central
star at the end of the simulations, but some are also bound to the planet, as
planetary satellites. The ``unbound'' population of planetesimals are those
particles moving on their own independent heliocentric orbits.

In the simulations, the bound and unbound populations are differentiated
according to two conditions.  We define the specific energy of a planetesimal
with respect to the planet, $E_{\rm rel}$,
\begin{equation}
E_{\rm rel} = - {GM_p \over r} + {1 \over 2}\;\left(\mathbf{v}-\mathbf{V_p}\right)^2\;,
\label{erel}
\end{equation}
where $r$ is the distance between the planet and the planetesimal,
$\mathbf{v}$ and $\mathbf{V_p}$ are the planetesimal and the planet
velocities, respectively (note that these quantities are calculated here at
$t=10$ whereas in \S \ref{sec:explosion} we measured relative positions and
velocities of planetesimals and the planet at $t=0$). The planetesimal is
considered unbound if
\begin{equation}
E_{\rm rel} > 0\;,
\label{ebound}
\end{equation}
and bound if 
\begin{equation}
E_{\rm rel} < 0\quad {\rm and} \quad r \le r_h'\;,
\label{unbound}
\end{equation}
where $r_h' = R_0 (M_p/3M_*)^{1/3}$ is the Hill's radius of the solid
planet. 

The two populations should be analysed differently, of course. The orbits of
the unbound population should be defined with respect to the central star,
whereas the orbits of the bound planetesimals are best defined with respect to
the planet. In the rest of this section we consider the unbound part of the
planetesimals only.

\subsection{Simulation U15}\label{sec:u15_unbind}

\begin{figure}
\centerline{
\psfig{file=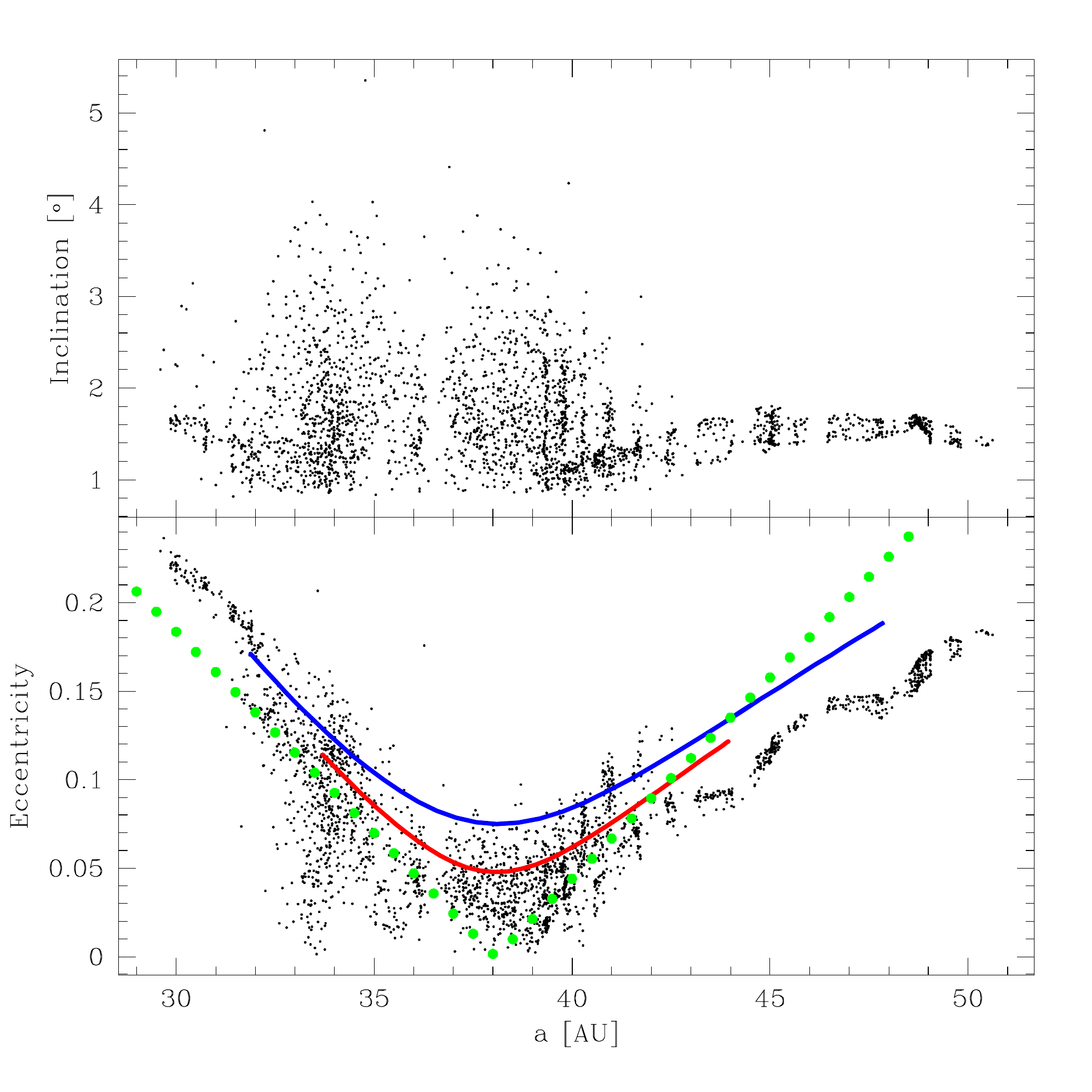,width=0.49\textwidth,angle=0}}
\caption{The inclination (top panel) and eccentricity (bottom panel) of the
  unbound planetesimals versus semi-major axis, $a$, in the simulation U15 at
  the end of the simulation, $t=10$. Individual planetesimals' orbital
  elements are shown with black dots. The coloured chevron-shaped curves are
  theoretical predictions (see \S \ref{sec:u15_unbind}). The nearly empty
  vertical contains many planetesimals still bound to the planet, which are
  not shown here but analysed in \S \ref{sec:bound_orbits}.}
\label{fig:EB015_orbit_LIVE}
\end{figure}

Fig. \ref{fig:EB015_orbit_LIVE} shows the orbital eccentricity (bottom panel)
and the inclination (top panel) of the unbound planetesimals versus the
semi-major axis of their orbits ($a$) at $t=10$. As explained above, their
orbital parameters are calculated with respect to the star as their orbits are
heliocentric.

The top panel of figure \ref{fig:EB015_orbit_LIVE} shows that planetesimals
continue to have very small orbital inclinations after the disruption, with
the mean inclination angle of only $i\approx 1^\circ$. This is natural since
in our setup tidal disruption of the host gas protoplanet is symmetric around
its orbital plane. In fact, the approximate analytical theory of \S
\ref{sec:explosion} predicts that inclination of planetesimal orbits should
remain exactly zero. Therefore, the inclination found in the simulation must
result either from a small but finite asymmetry with respect to $z \rightarrow
-z$ inversion developing during the disruption, or due to effects not taken
into account in the analytical theory.

The central $a\approx 32$ to $a\approx 40$ AU region of this plot shows a
higher dispersion in the values of $i$ than do the more distant regions. To
understand the origin of this result, we note that the central region is the
one where most of the gas ends up after being disrupted, e.g., see the right
panel of figure \ref{fig:OFF00EB015VY_late}. We also note that gas from the
disrupted host clump is initially arranged in a high density spiral,
gravitational potential of which may well be significant enough to scatter
planetesimals about and to pump their inclinations.

This explanation is consistent with the fact that planetesimals on wider
orbits, e.g., $a \simgt 42$ AU, which should be affected by the interactions
with the gas spiral less, indeed have smaller dispersion in their inclination
angles. The non-zero mean value of $i$ for this population, more distant from
the planet, should be due to asymmetries developing during the disruption
process. Comparison with simulation U0 below, which produces far smaller
inclinations, demonstrates that the asymmetries are probably related to the
injection of energy in the centre of the host in U15 (and U10). This energy
injection increases the entropy of the gas within $r < r_{\rm ej}$ around the
planet but not at larger radii. The initial entropy profile is thus strongly
unstable to convection in U10 and U15, most likely leading to an asymmetric
(with respect to the initial orbital plane of the clump) expansion of the host
clump.

The bottom panel of figure \ref{fig:EB015_orbit_LIVE} shows the planetesimals
in the eccentricity -- semi-major axis ($e$--$a$) plane.  The coloured curves
show the theoretically calculated correlations of $e$ and $a$ with slightly
different assumptions (explained below) on the basis of the approximate theory
of \S \ref{sec:theory}. All of these curves were shifted to $a_0 = 38$ AU to
fit the simulations better.  The shift may be empirically justified as
following. Our approximate theory is based on the assumption of an
instantaneous disruption of the protoplanet. However, in the simulations the
protoplanet is not disrupted instantaneously, and it does migrate inwards
during the disruption, presumably due to gas gravitational torques. Therefore,
in terms of our approximate theory, the ``effective'' location of protoplanet
disruption should be smaller than the initial radius of the host clump's
circular orbit.

Amongst the theoretically computed curves, the green dots show the simplest
one -- equation \ref{ecc_a}. We recall that this equation was obtained with
the help of a Taylor decomposition of the specific energy and the specific
angular momentum in powers of relatively small parameters $r/R$ and $\delta
v/V$. In contrast, the red and the blue curves show the predicted orbits of
the planetesimals that did not use the decomposition. To draw the curves, we
consider a ring of planetesimals of radius $r=0.3$ and $0.4$ AU for the red
and blue curves, respectively. We also assume that the kick velocity of the
planetesimals after the disruption, $\mathbf{\Delta v}$ is equal to $\delta
\mathbf{v}_{\rm circ}$, where $\delta = 0.5$ and $\mathbf{v}_{\rm circ}$ is
the circular velocity of the planetesimal before the disruption (cf. equation
\ref{vcirc}). This simple (but somewhat more accurate than the green dots)
theory also predicts the ``V'' shaped $e-a$ correlation.

We also note that our neglect of the second order terms in equations
\ref{deltae_app} and \ref{deltaL1} lead to the equation \ref{ecc_a} being
symmetrical with respect to the sign of $a-a_0$. The more exact calculations
given with the red and the blue curves are not quite symmetrical and also show
non-zero eccentricity for all the particles.  On the other hand, SPH/N-body
simulations show a wider distribution of eccentricities at a given $a$, with
some particles reaching $e\approx 0$.  Nevertheless, it appears that by
varying the parameter $\delta$ (which is only constrained to be less than
unity) and by considering the different rings of planetesimals $r$ within the
host before its disruption, we can qualitatively explain the observed range of
$e$ and $a$ as well as their correlation.

The approximate analytical theory of \S \ref{sec:theory} also makes a
prediction on the width of the disruption ring, as $w = 0.5 \zeta$, where
$\zeta$ is expected to be of order unity. The debris ring in simulation U15
has width of $\sim 20$ AU, commensurate with $\zeta =1 $. The simple
``instantaneous'' disruption model is thus reasonably accurate in predicting
the radial extent of the debris disc in this simulation.

\subsection{Unbound debris in U10}\label{sec:unbound_u10}

\begin{figure}
\centerline{
\psfig{file=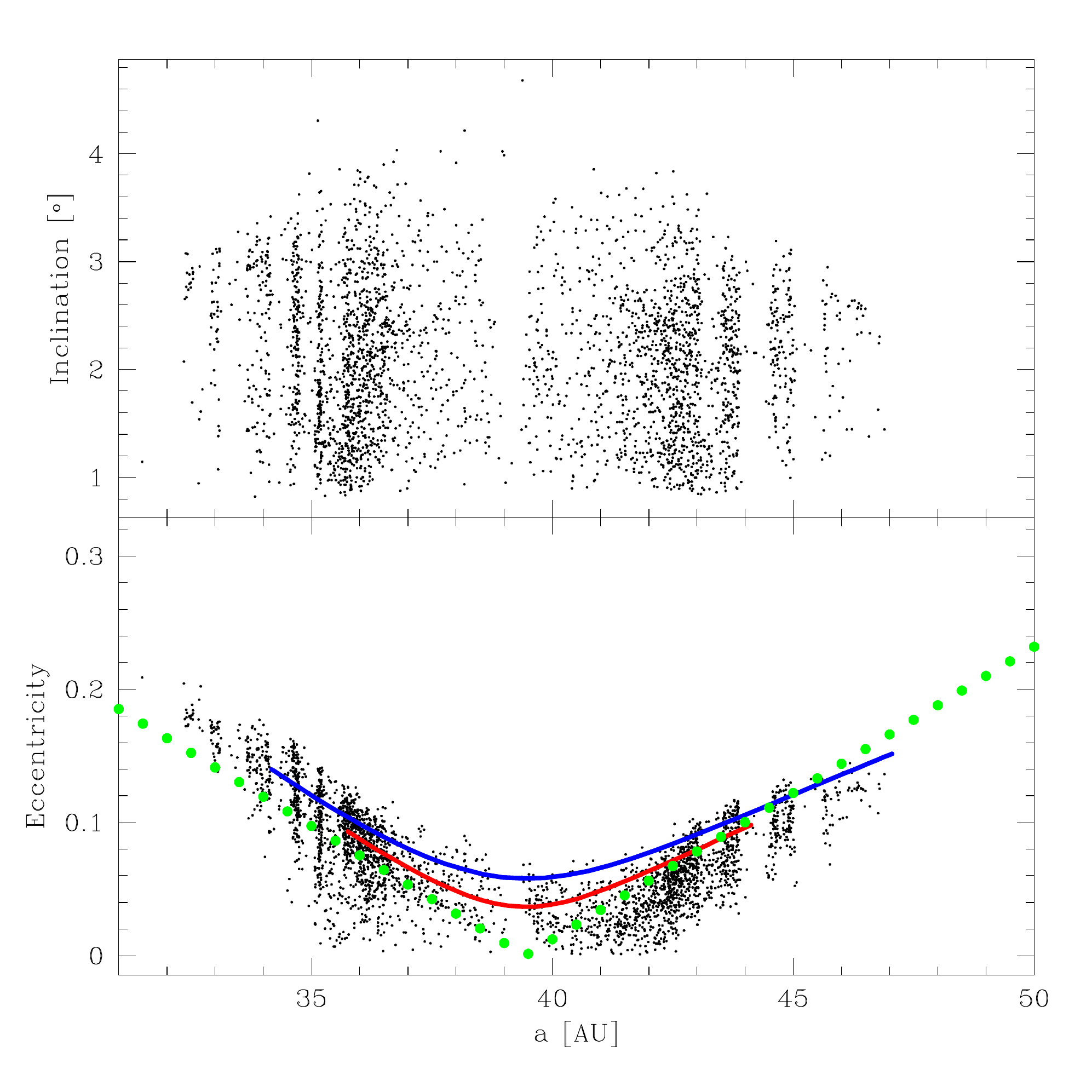,width=0.49\textwidth,angle=0}}
\caption{Same as Figure \ref{fig:EB015_orbit_LIVE} but at the end of
  simulation U10.}
\label{fig:OFF00EB010VY_RESUM_LIVE}
\end{figure}

Figure \ref{fig:OFF00EB010VY_RESUM_LIVE} shows the inclination and the
eccentricity of the unbound planetesimals in simulation U10, along with
theoretical curves, in the same format as in
fig. \ref{fig:EB015_orbit_LIVE}. Simulation U10 differs from U15 by the
smaller amount of internal energy injected into the host clump by the solid
core (planet). As the result, the disruption process is much more gradual and
protracted in U10 than in U15 (see \S \ref{sec:gas}).

The $e$--$a$ diagram shows a great deal of similarity between simulation U10
and U15, but also some interesting differences. For example, the radial width
of the disruption ring is narrower in the former. This makes intuitive sense
as there is less internal energy input into the gas clump in U10, and
therefore the velocity ``kicks'' $\Delta v$ after the disruption are probably
smaller than they are in U15.

The inclination angle $i$ versus $a$ plot shows somewhat larger mean value and
larger dispersion of $i$ in U10 than in U15. In addition, there are pronounced
vertical bands in the top panel of Figure
\ref{fig:OFF00EB010VY_RESUM_LIVE}. Finally, if we break the pattern of
fig. \ref{fig:EB015_orbit_LIVE} on the ``gas-dominated'' central and the
``far-out'' population at $a > 43$ AU, then its fair to say that in simulation
U10 the far-out population is all turned into the central gas-dominated one.

All of these differences are explainable by the fact that the disruption
process is more gradual in U10, which means that the tidal tails (arms) of the
host clump in the process of disruption (cf. the left panel of
fig. \ref{fig:OFF00EB010VY_3P_snap}) survive for longer, before being
circularised into a diffuse gaseous ring (the right panel of
fig. \ref{fig:OFF00EB010VY_3P_snap}). This implies that the planetesimals are
influenced stronger by these tidal tails, explaining a higher dispersion in
inclination $i$. In addition to that, the tidal tails ``shepherd'' the
planetesimals out of the cloud, as in seen in the left panels of figures
\ref{fig:OFF00EB015VY_late} and \ref{fig:BIG_snap}, causing bunching of the
orbital parameters in the vertical bands evident in the top panel of figure
\ref{fig:OFF00EB010VY_RESUM_LIVE}. As the spiral arms survive for longer in
U10 than they do in U15, the bunching effect is far stronger in the former than
in the latter.

\subsection{Unbound planetesimals in simulation U0}\label{sec:unbound_u0}

\begin{figure}
\centerline{
\psfig{file=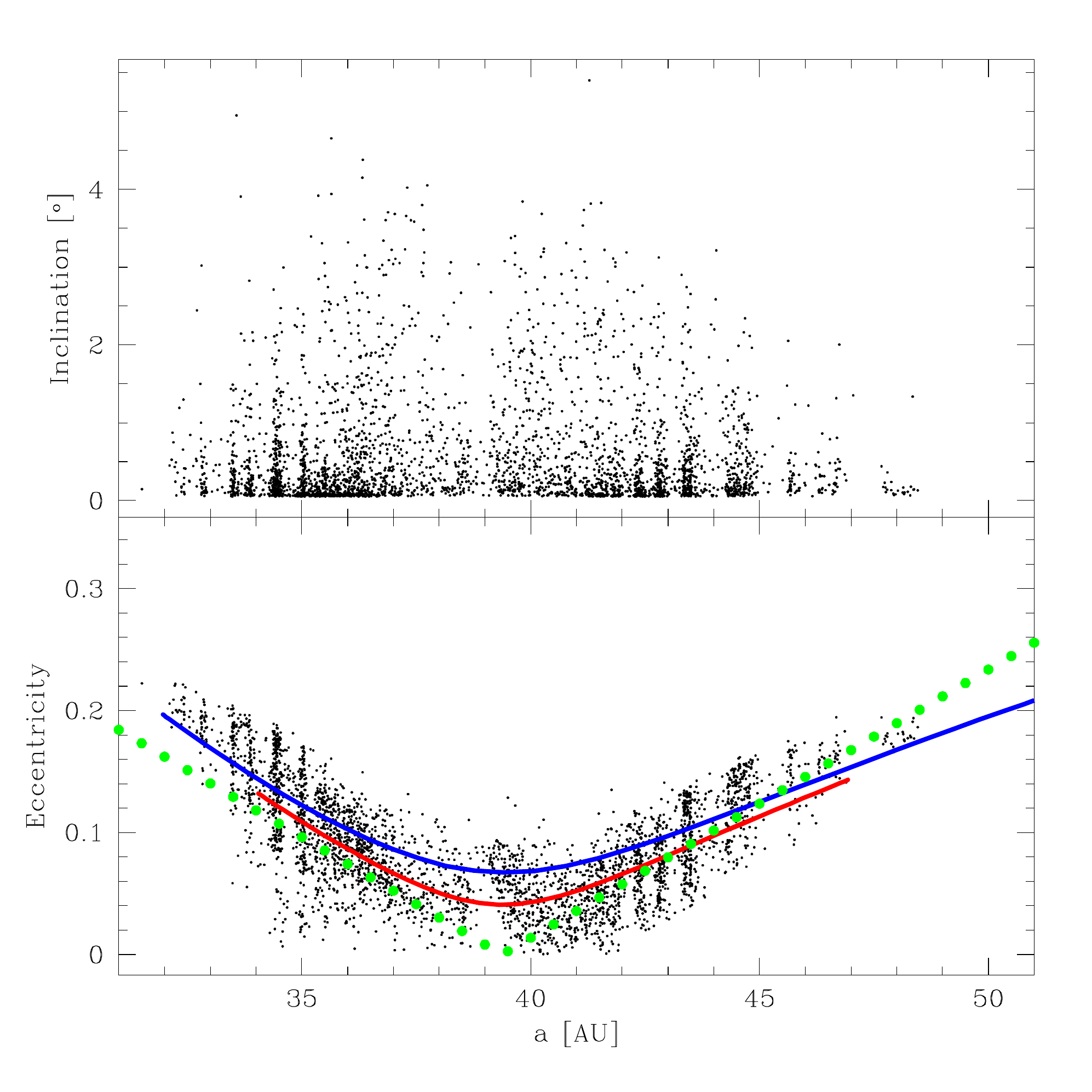,width=0.49\textwidth,angle=0}}
\caption{Same as Figs. \ref{fig:EB015_orbit_LIVE} and
  \ref{fig:OFF00EB010VY_RESUM_LIVE}, except for the red and blue curves, as
  explained in the text (see \S \ref{sec:unbound_u0}). Note the much larger
  $i\approx 0$ population of planetesimals in this simulation compared to U10
  and U15}
\label{fig:BIG_ANA}
\end{figure}

Figure \ref{fig:BIG_ANA} shows the orbital parameters of the unbound
planetesimals at the end of simulation U0. This figure shows, yet again, the
familiar $e$--$a$ correlation diagram but a significantly different
distribution of planetesimal orbital inclinations. The red and the blue curves
in this case were computed assuming two rings of planetesimals with $r=0.75$
and $1$ AU, respectively, because the planetesimal disc is larger in U0 than
in U10 and U15, and we used $\delta = 0.9$ (see \S \ref{sec:u15_unbind}). The
latter is chosen by trial and error to find a visually reasonable match to the
distribution of orbits in the simulation.

The much stronger bunching of planetesimals towards the $i=0$ plane is a
testament to the different nature of the host clump disruption in U0 as
compared to U10 and U15. While in the latter two the inner region of the clump
was the actual source of the disruption (as the core accretion energy was
dumped there), in U0 the innermost region is ``passive''. Therefore the
disruption process, as experienced by the planetesimals initially located at
$r \le 1$ AU for this simulation, is much less abrupt, leading to even less
inclined orbits for the unbound particles. In addition, as mentioned in
\ref{sec:u15_unbind}, the injection of energy stirs up strong convective
motions in those simulations, pumping up asymmetries and thus orbital
inclinations. 

On the other hand, the larger dispersion in values of $i$ in the centre of the
planetesimal ring, and the presence of the vertical bands in
fig. \ref{fig:BIG_ANA} confirms that these features are formed by the tidal
arms (tails) of the host clump before they are wound up and completely erased.

\section{Bound population: planet satellites}\label{sec:bound_orbits} 

We now switch to discussing planetesimals bound to the solid planet, e.g.,
those that have a negative energy with respect to the planet and that are
within the planet's Hill's radius (see \S \ref{sec:unbound}).

\begin{figure*}
\centerline{
\psfig{file=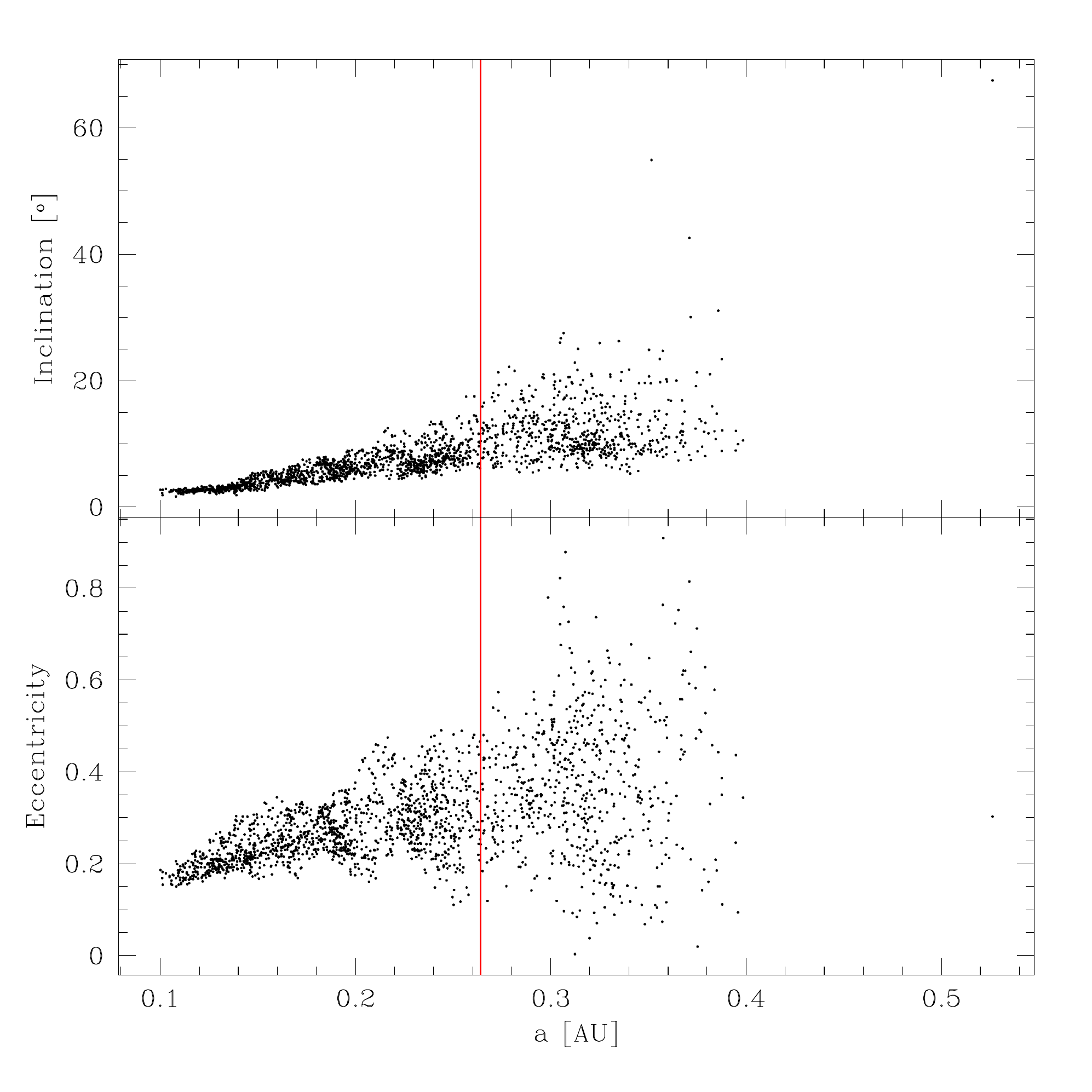,width=0.49\textwidth,angle=0}
\psfig{file=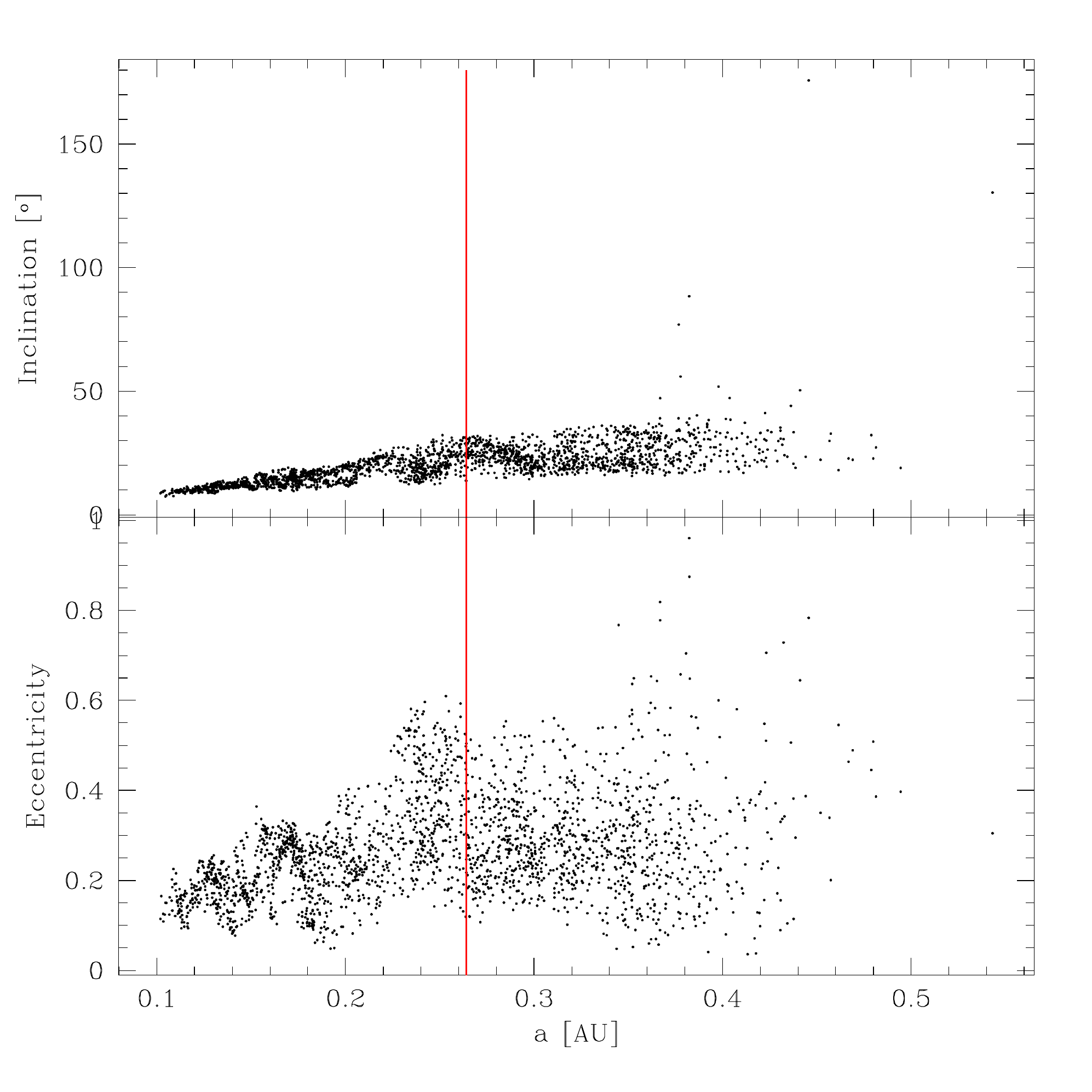,width=0.49\textwidth,angle=0}}
\caption{The bound population of planetesimals around the solid core
  (planet). The orbital characteristic are calculated with respect to the
  planet rather than the star. Left: Simulation U15, Right:
  simulation U10.}
\label{fig:U15_U10_sat}
\end{figure*}

Figure \ref{fig:U15_U10_sat} presents the orbital parameters of the
planetesimal orbits around the planet for the simulation U15 (left panel) and
U10 (right panel). The red vertical lines show the location of the planet's
influence radius, $r_i$, defined by equation \ref{ri}. This radius divides the
region where gravitational potential is dominated by the solid planet (inside
$r_i$) and the gas (outside $r_i$). As stated in \S \ref{sec:influence}, the
orbits of planetesimals are expected to be only mildly perturbed within $r_i$
and suffer strong perturbations outside $r_i$, in fact being completely
unbound from the planet at $r \gg r_i$.

Figure \ref{fig:U15_U10_sat} confirms this expectation qualitatively,
including the fact that there are very few particles with semi-major axis $a$
larger than 0.4 AU. The particles inside $r_i$ tend to have mild eccentricity
$0 \le e \le 0.4$, whereas planetesimals outside $r_i$ have higher values of
$e$ on average. The orbital inclination $i$ also increases with $a$, reaching
$10^\circ$ to $20^\circ$ outside $r_i$ in simulation U15, and slightly higher
values in U10. Note that a larger mean value of $i$ in U10 compared to that in
U15 found here for the bound population is consistent with a similar trend
that we found for the unbound population (cf. figures
\ref{fig:EB015_orbit_LIVE} and \ref{fig:OFF00EB010VY_RESUM_LIVE}).

\begin{figure}
\centerline{
\psfig{file=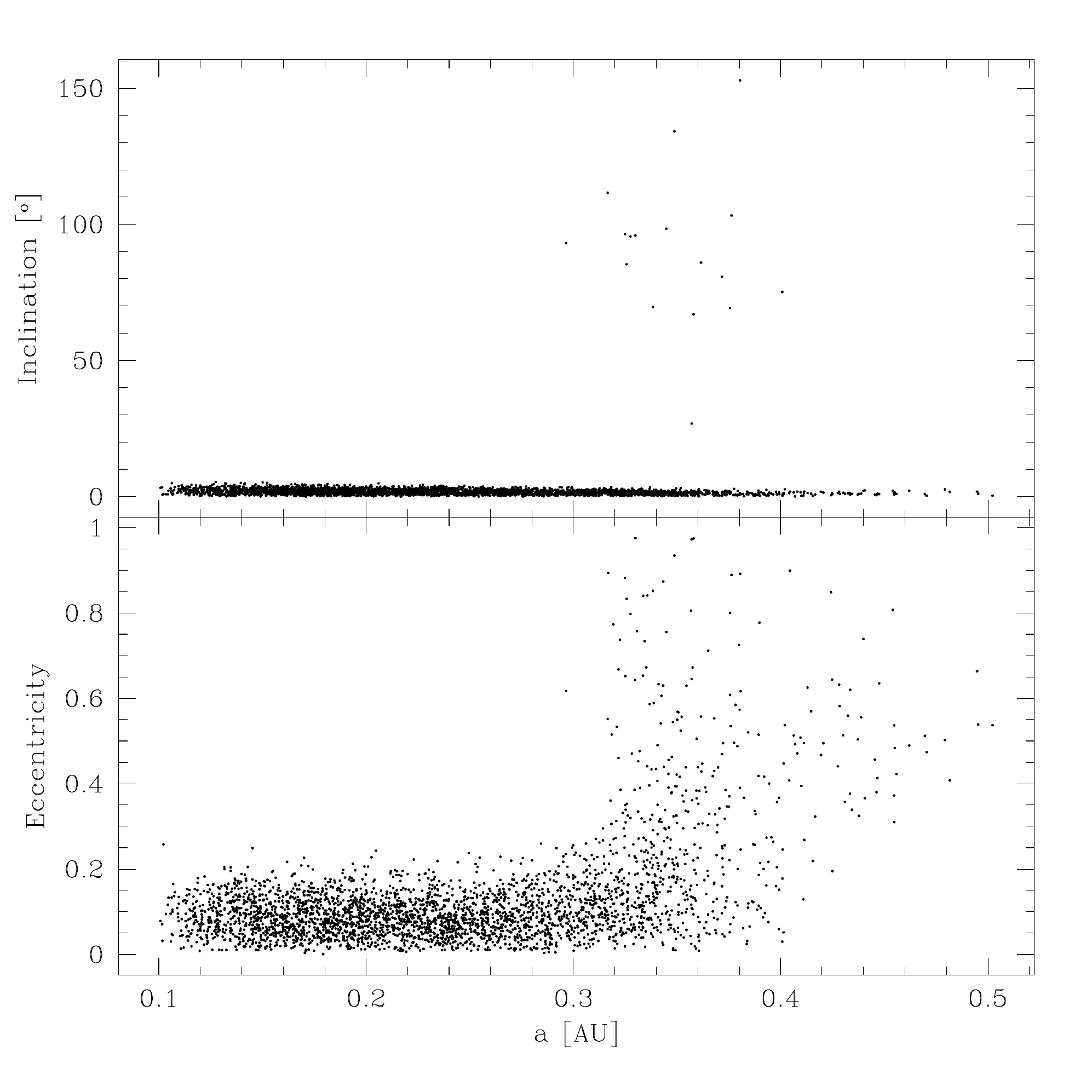,width=0.49\textwidth,angle=0}}
\caption{Same as Fig. \ref{fig:U15_U10_sat} but for simulation U0.}
\label{fig:U0_sat}
\end{figure}

Figure \ref{fig:U0_sat} shows the orbital parameters of the bound population
of planetesimals in simulation U0. There is a marked difference in these
distributions compared with that for U15 and U10. Instead of a monotonic
increase in eccentricities and inclinations with increasing $a$, evident in
Figure \ref{fig:U15_U10_sat}, here there is an almost a step-function change
in the nature of the orbits. Orbits at $a \simlt 0.3$ have small eccentricity,
$0 \le e \le 0.2$, whereas orbits at $a \simgt 0.3$ have a large spread in
eccentricities, with some approaching unity. The planetesimals with the
largest values of $e$ may become unbound later since their orbits are
comparable to $r_h$. There is also a group of particles with large
inclinations, $i\simgt 90^\circ$ at $a \approx 0.35$ AU.

The significant differences in the orbits of the bound populations of
planetesimals between Figures \ref{fig:U15_U10_sat} and \ref{fig:U0_sat}
demonstrate that the exact way in which the innermost region of the host clump
is disrupted influences the orbits of the ``satellites'' remaining after the
disruption. By extension it also means that the results would also be somewhat
different if there was a massive gas atmosphere around the solid
protoplanetary core.

\section{Discussion}\label{sec:discussion}

\subsection{General conclusions}\label{sec:general}

In this paper we considered the predictions of the TD hypothesis for the
``planetesimal'' debris left over after the disruption of a single gas host
clump. These gas clumps are the parent bodies within which all (or at least
most) of planet formation action takes place in the TD hypothesis.  The most
important results of our paper can be summarised as following:

1. In the context of TD hypothesis, large solids form only inside the massive
gas host clumps \citep[because the densities of solids initially reflect that
  of gas, and gas densities are orders of magnitude higher inside the host
  clumps than in the ``ambient disc''; see,
  e.g.,][]{Nayakshin10b,ChaNayakshin11a} that are then disrupted. The solid
debris is then spread around in a ring around the disruption location. TD
hypothesis thus generically predicts rings of planetesimals remaining
after planet formation rather than continuous discs.

2. The role of planetesimals in the formation of planets may be very
different in the CA and TD scenarios. Whereas planets could not have formed
without planetesimals forming first in the CA picture, in the TD hypothesis
this is less clear. In the latter case, once the initial proto solid core
  is born by gravitational collapse of the grain-dominated inner region of the
  host clump\citep{Nayakshin10a,Nayakshin10b}, further growth may be dominated
  by the proto solid core accreting $\sim$ cm to tens of cm grains. Solids in
this size range experience aerodynamic drag that is strong enough to dump
possible centrifugal support and yet small enough to sediment to the centre of
the host clump within its lifetime (see Fig. 1). Bodies in the planetesimal
size range, i.e., km-sized and larger, on the other hand, experience too weak
an aerodynamic drag. They are not expected to join the solid core en masse if
they have some orbital support due to an excess angular momentum or chaotic
convective motions of the gas in the centre of the host clump. In this case
planetesimals are no parent bodies of planets in the strict sense.

On the other hand, if turbulence, convection or angular momentum prevents
  small grains from reaching the proto solid core directly, turning most of
  these grains into planetesimals as discussed in \S \ref{sec:birth}, then the
  situation is less clear. If this happens sufficiently close to the proto
  solid core and planetesimal densities are high, most of that material can be
  subsequently accreted by the proto core somewhat like in the CA theory. In
  this case planetesimals are also building blocks of planets, with only those
  on larger less bound orbits managing to avoid being accreted onto the core.

3. Related to the point above, the solid planetary cores and the planetesimal
debris form at approximately same time in our picture. After host gas clump
disruption, some of the solid debris around the planetary core could fall on
it and be accreted in direct collisions, so that planetary growth could
actually continue in a manner analogous to that of the later oligarchic
stages of solid core assembly in the traditional scenario
\citep[e.g.,][]{Safronov72}. In particular, as velocity dispersion of the
  disrupted population is high, e.g. a fraction of a km~s$^{-1}$, only large
  planetesimals of size $\simgt 1000$ km would continue their growth, whereas
  smaller objects would accrete on large bodies or fragment in collisions with
  smaller bodies. If they fragment to sizes as small as a few metres then
  their inward radial migration inside the gas disc becomes important.

4. Also related to point (2) above, the mass budget of planetesimals can be
potentially very different in the two planet formation theories. In the CA
scenario, the initial population of planetesimals must have had a total mass
larger than the total high-Z element (other than H and He) mass of the
resulting planets. This does not {\em have to be} the case in the TD
scenario. In particular, there does not seem anything wrong with assuming that
the debris population may be far less massive than the protoplanetary core.

5. In the CA scenario, the properties of the protoplanetary disc are expected
to be a rather smoothly varying function of radius, except for the regions
near the ice line \citep[e.g.,][]{Armitage10}. Thus one expects that
planetesimal properties are also a smoothly varying function of $R$. In
contrast, in the TD hypothesis, it is the host clump that determines the
outcome of the planet formation process. The internal structure of the clump
is a strong function of its mass \citep{Nayakshin10a,Nayakshin10b}; that structure is
non-uniquely coupled to radius $R$ where the clump is disrupted. It is
possible that clumps of different history, mass, internal structure are
disrupted at roughly same location. Further, we envisage a number of host
clumps forming and being destroyed during the early gas-rich phase as required
to explain the FU Ori outbursts of young protostars
\citep{NayakshinLodato11}. This implies a far greater diversity in the
properties of the solid debris at the same spatial location than is possible
in the CA model.

6. We found in this paper that disruption of a host gas clump generically
produces a ``V''-shaped pattern in the eccentricity-semi-major axis space for
the planetesimal population released by the disruption. The planetesimals in
the centre of the pattern have nearly circular orbits while whose at the edges
of it are much more eccentric. The maximum eccentricities possible can be
estimated based on the simple analytical theory; from equations \ref{ecc_a} to
\ref{w_predict}, it follows that, within a factor of $\sim 2$, it is $e_{\rm
  max} \sim r_h/R \approx 0.2$ for host clump mass of $\sim 10$ Jupiter
masses.

7. The eccentricities and inclinations of the disrupted planetesimal
population are generally much too high to allow planetesimals to stick to one
another. The typical dispersion velocity of the disrupted population is
$\delta v \sim v_K r_h/R \sim 0.1 v_K$. At the distance of the Kuiper belt,
for example, this yields dispersion velocity of the order of $300$ m/s. Only
Pluto-sized bodies could survive {\em equal-size} collisions at these
velocities. Accretion on massive solid cores, however, is allowed as already
noted in (3) above, so the post-disruption evolution may be somewhat similar
to the final phases of the run-away growth of planetary embryos in the
traditional scenario \citep[e.g.,][]{Safronov72}.


8. We also find that solid bodies on orbits tightly bound to the planetary
core could survive the disruption of the gas host clump, remaining bound to
the planet. These bodies are satellites of the planetary cores.
  Satellites on orbits more strongly bound to the planet are affected by the
  gas envelope destruction less than those on less bound orbits. Some of the
  outermost satellites may even find themselves in eccentric counter-rotating
  orbits (e.g., see fig. \ref{fig:U15_U10_sat}). This is qualitatively similar
  to the satellites of the giant planets in the Solar System, although we note
  that our simulations are not designed to study these issues; one also would
  need to model composition, size differences and long-term survivability of
  the bound objects.

\subsection{Potential relevance to the Solar System}\label{sec:ss}

Even though we did not specifically attempt to reproduce the structure of the
outer Solar System here (which would require an additional study of how the
system evolves on the 4.5 Giga year time scale), our simulations do have
potentially interesting implications for it. We mention these implications
only briefly here, with the intent of quantifying them in a future
publication:

\begin{itemize}

\item One of the key properties of the trans-Neptunian region of the Solar
  System, including the Kuiper belt, is the ``mass deficit problem''. The
  current mass of the Kuiper belt is estimated at $\sim 0.01 - 0.1 \mearth$
  \citep{BernsteinEtal04}. On the other hand, to grow the observed populations
  of solid bodies in the context of the \cite{Safronov72} model for formation
  of solids, 10 to 100 $\mearth$ of solids is required
  \citep[e.g.,][]{KL99}. Thus, removal of $\simgt 99.9$\% of solid material is
  required. In the TD hypothesis, solids are made within the central fractions
  of AU of the host gas clumps \citep{Nayakshin10a}. The host clumps contain
  as much as $\sim 60\mearth$ of high-Z elements \citep{Nayakshin10b} which
  are then partially used to make the massive solid cores, the
  ``planetesimals'', and partially remain bound to gas in small grains. If it
  is possible to build solid cores as massive as $\sim 10 \mearth$ by direct
  gravitational collapse and accretion of $\sim$cm size grains
  \citep{Nayakshin10b}, it should also be possible to build Pluto-sized
  objects {\em without} having to put tens of $\mearth$ of material into the
  planetesimals. There is thus no mass deficit problem for the Kuiper belt in
  the TD scenario.

\item There is a sharp outer edge to the Kuiper belt at around $R\approx 50$
  AU which is currently not well understood \citep[e.g.,][]{Morbidelli08}. The
  disruption of a gas host clump naturally produces a ring with sharp outer
  and inner edges (point 1 above) because the range of orbits available to
  planetesimals after disruption is limited by the parameters of the initial
  host clump. This could naturally account for the outer edge of the Kuiper
  belt objects.

\item The structure of the classical Kuiper belt is best explained by our
  model if we assume that the left hand side (lower $a$) of the disruption
  ring (e.g., Fig. \ref{fig:BIG_ANA}) has been destroyed by interactions with
  planets. Realistically, the planet, left at $a\sim 35-39$ AU at the end of
  our simulations, could have continued to migrate inward via type I migration
  in the gas disc: the disrupted gas ring is still quite massive (5 Jupiter
  masses for our simulations here, and there would be more at larger $R$ to
  effect the migration of the host clump in the first place). Using the
  standard type-I gas migration \citep[e.g.,][]{Tanaka02} for Neptune
  initially located at $\approx 40$ AU, assuming the disc mass being $\sim 10
  $ Jupiter masses, and the gas disc aspect ratio of $H/R \sim 0.1$ yields
  type I migration rate time scale of $\sim 10^6$ yrs. This is sufficiently
  fast to allow Neptune to migrate inward significantly to end up, for
  example, in the configuration proposed by the NICE model of the outer Solar
  System \citep{GomesEtal05}. While migrating inward in the gas disc, Neptune
  would have scattered the left part of the ``V''-pattern of the
  planetesimals, but leaving behind the right hand side of the pattern as a
  belt reminiscent of the Kuiper belt.

\item Kuiper belt contains the hot and the cold populations. These are best
  accounted for by two different gas clump disruptions in our model. The
  hot population in our model would have most naturally resulted if the host
  gas protoplanet rotation axis were highly inclined to the orbital plane. In
  this case one could end up with higher inclinations than we obtained here.

\item \cite{NesvornyYR10} show that if planetesimals form by a local
  gravitational collapse in a high density environment, then one can naturally
  explain the surprisingly large fraction of binaries in the $\sim $100 km
  class low-inclination objects in the classical Kuiper Belt. We note that our
  model produces ``planetesimals'' in a similar fashion (gravitational
  collapse in a high density environment) albeit inside of the host clumps
  rather than the disc. Therefore by extension one might expect to see a high
  fraction of massive objects locked in binaries after the dissipation of the
  host clump. This idea however needs to be checked with a longer time
  $N$-body calculation as there are non trivial physical constraints on
  collisional destruction of binaries in the Kuiper Belt
  \citep{NesvornyEtal11}.

\end{itemize}

\section{Conclusions}

We have considered the origin of solid debris, such as comets, asteroids and
large bodies such as Pluto in the context of a recent planet formation
hypothesis (Tidal Downsizing). We assumed (see \S \ref{sec:birth}) that
the solid ``debris'' is formed in a way very similar to the massive solid
protoplanetary cores themselves, e.g., inside the $\sim 10$ Jupiter mass host
gas clouds. The latter are eventually disrupted either tidally or due to an
internal energy release during the solid core accretion. The release of these
solid bodies into the field forms rings potentially reminiscent of the Kuiper
belt and the debris discs around nearby main sequence stars. While much work
remains to be done to detail predictions of the TD hypothesis further, it is
already clear that these predictions are sufficiently different from the
standard planetesimal-based paradigm for planet formation \citep{Safronov72}
to be critically tested by observations in the near future.

As an astrophysical aside, we note that the rapid accretion of large solid
bodies (Pluto-like and even Neptune-like) in the TD scheme suggests that
planet and solid debris formation is a very robust process and may even occur
in very crowded environments such as the inner parsecs of galaxies
\citep{NayakshinSS12,ZubovasNM12}; this is an untenable proposition in the CA
model.

\section*{Acknowledgments}

Theoretical Astrophysics research in Leicester is supported by an STFC rolling
grant. This research used the ALICE High Performance Computing Facility at the
University of Leicester. Some resources on ALICE form part of the DiRAC
Facility jointly funded by STFC and the Large Facilities Capital Fund of
BIS. We acknowledge constructive report by the anonymous referee whose
suggestions helped to improve the clarity of this paper significantly.



\label{lastpage}

\end{document}